\documentclass[12pt]{article}
\usepackage[T1]{fontenc}
\usepackage{a4wide}
\usepackage{amsmath}
\usepackage{amsfonts}
\usepackage{amssymb}
\usepackage[dvips]{graphics}
\usepackage{psfrag}
\usepackage{subfigure}

\makeatletter
\newcommand{\ud}{{\mathrm d}}

\renewcommand\theequation{\hbox{\normalsize\arabic{section}.\arabic{equation}}}
\@addtoreset{equation}{section}

\@addtoreset{figure}{section}

\@addtoreset{table}{section}
\makeatother

\begin{document}

\title{
\begin{flushright}
\normalsize{
{ITP--Budapest Report No. 559\\
KCL--MTH--00--50
}}
\end{flushright}
\vspace{1cm}
\textbf{Nonperturbative study of the two-frequency sine-Gordon model}
}
\author{Z. Bajnok$ ^{1}$\thanks{
bajnok@poe.elte.hu
}, L. Palla$ ^{1}$\thanks{
palla@ludens.elte.hu
}, G. Tak{\'a}cs$ ^{2}$\thanks{
takacs@mth.kcl.ac.uk
} and F. W{\'a}gner$ ^{1}$\thanks{
wferi@afavant.elte.hu
}\\
\\
$ ^{1}$ Institute for Theoretical Physics, E{\"o}tv{\"o}s
University\\
H-1117 Budapest, P{\'a}zm{\'a}ny P{\'e}ter s{\'e}t{\'a}ny 1/A\\
Hungary\\
\\
$ ^{2}$ Department of Mathematics, King's College London\\
Strand, London WC2R 2LS, UK}

\maketitle
\begin{abstract}
The two-frequency sine-Gordon model is examined. The focus is mainly
on the case when the ratio of the frequencies is $1/2$, given the
recent interest in the literature. We discuss the model both in a
perturbative (form factor perturbation theory) and a nonperturbative
(truncated conformal space approach) framework, and give particular
attention to a phase transition conjectured earlier by Delfino and
Mussardo. We give substantial evidence that the transition is of
second order and that it is in the Ising universality
class. Furthermore, we check the UV-IR operator correspondence and
conjecture the phase diagram of the theory.
\end{abstract}

{\par\centering PACS codes: 64.60.Fr, 11.10.Kk \\[.5cm] 
}
 
\clearpage

\section{Introduction}

The sine-Gordon model has attracted a great deal of interest over the
decades, for the particular reason that while being an integrable
field theory, it can be used as a toy model for many nonperturbative
quantum field theory phenomena. It also has very interesting
applications in diverse areas of physics, ranging from statistical
mechanics of one-dimensional quantum spin chains to nonlinear optics
(for a -- non-exhaustive, but representative -- list with references
see the introduction of \cite{delfino_mussardo}).

In the present paper we study a nonintegrable extension of sine-Gordon
theory in which the scalar potential consists of two cosine terms with
different frequencies, called the two-frequency or double sine-Gordon
(DSG) model. It is suggested in \cite{delfino_mussardo} that this
model can be used as a more refined approximation to some of the
physical situations (e.g. wave propagation in a nonlinear medium)
where the ordinary sine-Gordon model is applicable. The model is also
interesting theoretically, since due to its nonintegrability it is
expected to have more general behaviour than usual sine-Gordon theory
and therefore it can be seen as a more realistic toy model for
nonperturbative quantum field theory. Applications to the study of
massive Schwinger model (two-dimensional quantum electrodynamics) and a
generalized Ashkin-Teller model (a quantum spin system) are
discussed in \cite{delfino_mussardo} and another potentially
interesting application to the one-dimensional Hubbard model is
examined in \cite{nersesyan} (together with the generalized
Ashkin-Teller model mentioned above). A further potentially interesting
application of the two-(and multi-)frequency sine-Gordon model is for
ultra-short optical pulses propagating in resonant degenerate medium
\cite{bullough}. 

We show that for a rational ratio of the frequencies this model is
still amenable to a nonperturbative treatment using the truncated
conformal space approach (TCSA), and that a great deal of interesting
information can be extracted, in particular the existence and the
characteristics of a phase transition which was predicted in
\cite{delfino_mussardo} on the basis of simple classical (mean field)
arguments.

The outline of the paper is as follows. In Section \ref{sec:basics} we
introduce the model and discuss some of its basic features. Then in
Section \ref{sec:tcsa} we set up the TCSA framework for the
model. Section \ref{sec:ffpt} discusses results obtained from the
so-called form factor perturbation theory (FFPT), while in Section
\ref{sec:comparison} we compare the FFPT results to the numerical data
resulting from TCSA in order to test the reliability of the latter
method. Section \ref{sec:phasetransitions} is devoted to a preliminary
discussion of phase transition in the case when the ratio of the two
frequencies is $1/2$ (denoted by $\text{DSG}_2$) and contains a
general analysis of signatures of second and first order phase
transitions in finite volume in a separate subsection. In Section
\ref{phasediagram} we apply these general considerations to construct
the phase diagram of the model $\text{DSG}_2$ and then we make our
conclusions in Section \ref{sec:conclusions}. The paper contains an
Appendix which collects the formulas used for computing the first
order corrections in form factor perturbation theory.

\section{The two-frequency sine-Gordon model}
\label{sec:basics}

\subsection{Definition of the model \label{subsec:definition}}

The two-frequency sine-Gordon (or double sine-Gordon, DSG) model is defined
by the action
\begin{equation}
\label{DSG_action}
\mathcal{A}_{\text{DSG}}=\int \ud t\int \ud x\left( 
\frac{1}{2}\partial _{\mu}\Phi \partial ^{\mu }\Phi 
+\mu \cos \beta \Phi +\lambda \cos \left( \alpha \Phi +\delta \right) \right) \,,
\end{equation}
where $ \beta  $, $ \alpha  $ are the two frequencies, $ \mu  $, $ \lambda  $
are dimensionful couplings (classically of dimension mass$ ^{2} $) and $ \delta  $
is a relative phase. Setting $ \mu  $ or $ \lambda  $ to zero the DSG
model reduces to an ordinary sine-Gordon model.

At the quantum level this theory can be considered as a double perturbation
of its $ c=1 $ UV limiting conformal field theory. The action can be written
in the following form
\begin{equation}
\mathcal{A}_{\text{DSG}}=\mathcal{A}_{c=1}-\mathcal{A}_{\text{pert}}\,,\end{equation}
where
\begin{eqnarray}
\mathcal{A}_{c=1} & = 
& \int \ud t\int \ud x\frac{1}{2}\partial _{\mu }\Phi \partial ^{\mu }\Phi\,, \\
\mathcal{A}_{\text{pert}} & = & -\frac{1}{2}\int \ud t\int \ud x\left( \mu
V_{\beta }+\mu V_{-\beta }
+\lambda e^{i\delta }V_{\alpha }+\lambda e^{-i\delta }V_{-\alpha }\right) \,,
\end{eqnarray}
and $ V_{\omega } $ denotes the exponential field (vertex operator)
\begin{equation}
V_{\omega }=e^{i\omega \Phi }\,,\end{equation}
which in the UV limit corresponds to a ($ \widehat{U(1)} $ Kac-Moody) primary
field with conformal dimensions 
\begin{equation}
\Delta ^{\pm }_{\omega }=\Delta _{\omega }=\frac{\omega ^{2}}{8\pi }\: ,\end{equation}
where the upper index $\pm$ corresponds to the right/left conformal
algebra. Correspondingly the couplings at the quantum level have the following dimensions
\begin{eqnarray}
\left[ \mu \right]  & = & \left( \text{mass}\right) ^{2-2\Delta _{\beta }}\: ,\\
\left[ \lambda \right]  & = & \left( \text{mass}\right) ^{2-2\Delta _{\alpha }}\: .
\end{eqnarray}
We also introduce two mass scales $ M_{\alpha } $ and $ M_{\beta } $,
which are defined to be the masses of the sine-Gordon solitons in the models with
$ \mu =0 $ or $ \lambda =0 $, respectively. It is known \cite{mass_scale}
that these masses are related to the couplings $ \lambda  $, $ \mu  $ via
\begin{eqnarray}
\mu  & = & \kappa _{\beta }M_{\beta }^{2-2\Delta _{\beta }}\: ,\nonumber \\
\lambda  & = & \kappa _{\alpha }M_{\alpha }^{2-2\Delta _{\alpha }}\,, \label{massgap_relation} 
\end{eqnarray}
where 
\begin{equation}
\label{kappa}
\kappa _{\omega }=\frac{2\Gamma (\Delta _{\omega })}{\pi \Gamma
(1-\Delta _{\omega })}
\left( \frac{\sqrt{\pi }\Gamma \left( \frac{1}{2-2\Delta _{\omega
}}\right)}{2 \Gamma \left( \frac{\Delta _{\omega }}{2-2\Delta _{\omega
}}\right) }\right) ^{2-2\Delta _{\omega }}\, .
\end{equation}

\subsection{General properties of the model}

The DSG model was examined in \cite{delfino_mussardo}
by Delfino and Mussardo, where some general observations were made. Here we
recall these facts without eventually going into any details. 

The first observation is a simple consequence of the formulation of the model
as a perturbed $ c=1 $ CFT. Namely, in order for the two perturbing operators
to be relevant, one has to restrict
\begin{equation}
\beta ^{2}\leq 8\pi \quad ,\quad \alpha ^{2}\leq 8\pi \, .\end{equation}
Furthermore, we  assume that the theory is renormalizable in the strict
sense, i.e. all divergences can be absorbed into redefining the coupling constants
$ \mu  $ and $ \lambda  $ (otherwise one needs to add new  (periodic) counterterms
to the action and the theory is no more a two-frequency sine-Gordon model but
rather some multi-frequency version). This means that we have to restrict
\begin{equation}
\alpha \beta \leq 4\pi \, .\end{equation}
In the sequel we  always assume that the above two conditions hold.

Another nontrivial property of the model is that it is a nonintegrable
field theory\footnote{One can argue for the nonintegrability
e.g. noting that the so-called ``counting argument'' fails to provide
any conserved higher spin quantities if $\alpha \neq \beta$ and both
$\lambda$ and $\mu$ are nonzero.} if $ \alpha \neq \beta $ and both
couplings $ \lambda $, $ \mu $ are nonzero. In addition, the model
exhibits drastically different behaviour depending on whether the
ratio of the two frequencies is rational or irrational.

\subsubsection{Rational case}

If  
\begin{equation}
\label{rational_ratio}
\frac{\alpha }{\beta }=\frac{m}{n}
\end{equation}
with $ m $ and $ n $ two relative prime positive integers, then the potential
$ -\mu \cos \beta \Phi -\lambda \cos \left( \alpha \Phi +\delta \right)  $
is periodic with period 
\begin{equation}
\frac{2\pi n}{\beta }=\frac{2\pi m}{\alpha }\,.\end{equation}
It can then be shown that the fundamental range of the $ \delta  $ parameter
is 
\begin{equation}
\label{delta_range}
|\delta |\leq \frac{\pi }{n}
\end{equation}
(i.e. any model with other values of the parameter $\delta$ can be
redefined into one with a $\delta$ lying in the fundamental range by
shifting the field with some integer multiple of $ \frac{2\pi }{\beta
} $).  The periodicity of the potential implies that the theory has
an infinitely degenerate ground state and supports topologically
charged excitations. The fundamental topological excitation
degenerates in the
$ \lambda =0 $ limit to an $ n $-soliton state of the
corresponding sine-Gordon theory and similarly for the $ \mu =0 $
limit it is an $ m
$-soliton state. For general values of the couplings, the solitons
are in a sense ``confined'' into these composite objects.

\subsubsection{Irrational case}

If, on the other hand, 
\begin{equation}
\frac{\alpha }{\beta }\notin {\mathbb Q}\,,\end{equation}
then the potential is not periodic. There are no topologically charged excitations
and the solitons are completely confined. Delfino and Mussardo \cite{delfino_mussardo}
argue that in this case the ground state is unique. Furthermore, from (\ref{delta_range})
we can infer that the parameter $ \delta  $ presumably plays no role in the
dynamics of the model.

The irrational case is extremely complicated, even as a quantum
mechanical problem (i.e. neglecting the $ x $ dependence). In the
next section we shall see that it is impossible to apply the
well-known TCSA method in this case and in lack of any alternative
nonperturbative method in the present paper we restrict our attention to the
rational case (although the perturbative formulas presented in Section
\ref{sec:ffpt} apply in the irrational case as well).

\section{Truncated Conformal Space Approach}
\label{sec:tcsa}

Here we only review the specifics of TCSA as applied to the DSG
model. This nonperturbative (numerical) method was invented by Yurov
and Zamolodchikov in \cite{yurov_zamolodchikov} and extended to $ c=1
$ theories in \cite{frt2} to which we refer the interested reader for
more details.

\subsection{The UV Hilbert space}

The first important issue is the operator content of the UV limiting $ c=1 $
CFT. We need a theory in which the perturbing operators $ V_{\pm \alpha } $,
$ V_{\pm \beta } $ correspond to local operators.

In the $ c=1 $ CFT we consider the spectrum of local operators is
generated by the conformal families of $ \widehat{U(1)} $ Kac-Moody
primary fields $ \mathcal{V}_{r,s} $ with scaling dimensions
\begin{equation}
\Delta _{r,s}^{\pm }=\frac{1}{2}\left( \frac{r}{R}\pm \frac{1}{2}sR\right) ^{2}\; ,\; r,s\in {\mathbb Z}\: ,\end{equation}
where $ R $ is the compactification radius of the $ c=1 $ free boson field
$ \chi  $ with the action
\begin{equation}
\mathcal{A}_{\chi }=\frac{1}{8\pi }\int \ud ^{2}x\partial _{\mu }\chi \partial ^{\mu }\chi \quad ,\quad \chi \sim \chi +2\pi R\: .\end{equation}
Taking into account eqn. (\ref{rational_ratio}) and the fact that
the normalization of $\Phi$ and $\chi$ differ by $\sqrt{4\pi}$, their
identification is essentially unique up to a positive integer $ k
$:
\begin{eqnarray}
 &  & R=k\frac{\sqrt{4\pi }n}{\beta }=k\frac{\sqrt{4\pi }m}{\alpha }\,,\\
 &  & V_{\pm \beta }\equiv \mathcal{V}_{\pm nk,0}\quad ,\quad V_{\pm \alpha }\equiv \mathcal{V}_{\pm mk,0}.
\end{eqnarray}
The integer $ k $ is the folding number, which has been considered in the
context of sine-Gordon theory in \cite{kfold} and can be introduced in any
scalar field theory with periodic potential: it corresponds to identifying the
field as an angular variable after $ k $ periods of the potential. We 
choose $ k=1 $, which means we identify
\begin{equation}
\label{angular_nfold}
\Phi \equiv \Phi +\frac{2\pi n}{\beta }l\quad ,\quad l\in {\mathbb Z}\: .
\end{equation}
However, it can now be seen that in the limit $ \lambda =0 $ we recover an
$ n $-folded version of sine-Gordon theory (denoted $ \text{SG}(\beta ,n) $
in \cite{kfold}) while in the $ \mu =0 $ case we get the theory $ \text{SG}(\alpha ,m) $.
We remark that by choosing $ k=1 $ we do not lose any information on the
DSG model, since the other $ k>1 $ folded versions can be obtained from the
$ k=1 $ case by adding some twisted sectors as in \cite{kfold}. However,
we shall see that the fact that the sine-Gordon limits are in general
{\it folded models} will play a very important role in the dynamics of
the DSG model. There is also a fermionic version of the theory which
has primary fields $ \mathcal{V}_{r,s} $ in the UV with
\begin{equation}
s\in {\mathbb Z}\: ,\: r\in {\mathbb Z}+\frac{s}{2}\end{equation} but
similarly to the folded ones, this version can be constructed by
replacing some of the topologically charged sectors with some
(fermionic) twisted ones. As we are only interested in the dynamics of
the vacuum sector we shall not discuss this modification here either.

We also wish to note that as the irrational case can be considered as a limit
$ n,m\, \rightarrow \, \infty  $, the Hilbert space of the irrational DSG
model corresponds to an uncompactified boson in the ultraviolet limit, with
a continuous spectrum of primary fields
\begin{equation}
\mathcal{V}_{r,0}\quad ,\quad \Delta _{r}^{\pm }=\frac{r^{2}}{2}\quad ,\quad r\in {\mathbb R}\: .\end{equation}
The fact that the spectrum is continuous means that TCSA is not applicable since
the conformal Hilbert space contains infinitely many states even when truncated
by imposing some upper bound on conformal energy.

\subsection{The TCSA Hamiltonian}\label{subsec:tcsa_ham}

We put the system on a cylinder with finite spatial volume $ L $,
i.e. we take the space coordinate $ 0\leq x<L $ and identify $x$ with
$x+L$. Given the identification (\ref{angular_nfold}) this means in
fact the following quasi-periodic boundary condition for the field $
\Phi $:
\begin{equation}
\Phi (x+L)=\Phi (x)+\frac{2\pi n}{\beta }q\quad ,\quad q\in {\mathbb
Z}\,.
\end{equation} 
$ q $ is the winding number labelling different
sectors of the theory, and corresponds to a conserved topological
charge in the full perturbed CFT.  In the present paper we are only
interested in the $ q=0 $ sector as it contains all the relevant
information for the problems treated in this work. In addition, in the sector
containing the vacuum the Lorentz spin of all the states is also zero,
i.e. $ \left( L_{0}-\bar{L}_{0}\right) \left| \Psi \right\rangle =0 $
for any state $ \left| \Psi \right\rangle $.

After a mapping onto the conformal plane, the Hamiltonian of the theory can
be written in the following form
\begin{equation}
H_{\text{DSG}}=\frac{2\pi }{L}\left(
L_{0}+\bar{L}_{0}-\frac{c}{12}+\frac{\mu L^{2-2\Delta _{\beta
}}}{2(2\pi )^{1-2\Delta _{\beta }}}\left( V_{\beta }(1)+V_{-\beta
}(1)\right) +\frac{\lambda L^{2-2\Delta _{\alpha }}}{2(2\pi
)^{1-2\Delta _{\alpha }}}\left( V_{\alpha }(1)+V_{-\alpha }(1)\right)
\right) 
\end{equation} 

The matrix elements of this operator can be calculated explicitly
between any two states in the UV limiting Hilbert space. Truncating
this space to finitely many state by introducing an upper conformal
energy cutoff the Hamiltonian becomes a finite matrix, which can then
be diagonalized numerically to get an approximation of the spectrum.
Thus one can extract the energy $E_{\Psi}(L)$ of any state $\Psi$ as
a function of the volume $L$.

It is well-known \cite{klassen_melzer2} that TCSA suffers from UV
divergences whenever the scaling dimension $\Delta^++\Delta^-$ of any of the
perturbing operators is larger than $1$. In this case the only
quantities that are possible to extract are the energies relative to
the ground state i.e. $E_i(L)-E_0(L)$. Furthermore, in
the case where TCSA is UV divergent the truncation errors are much
larger even for such relative energies, and grow very fast with the
volume $L$, which restricts the usefulness of TCSA in this case to 
small values of $L$.

\subsection{The interpolating mass scale}

In order to get numerical results, we have to fix the units in
which we measure energies and distances. For later convenience, we introduce
the following \textit{`interpolating' mass scale} 
\begin{equation}
\label{interpolating_mass_scale}
M=\frac{M_{\beta }\mu ^{x_{\alpha }}+M_{\alpha }\lambda ^{x_{\beta }}}{\mu ^{x_{\alpha }}+\lambda ^{x_{\beta }}}\,,
\end{equation}
where $ x_{\omega }=2-2\Delta _{\omega }. $ We  also use the following
dimensionless combination of the coupling constants
\begin{equation}
\eta =\frac{\lambda ^{x_{\beta }}}{\mu ^{x_{\alpha }}+\lambda
^{x_{\beta }}}\,.
\label{eta_definition}
\end{equation}
The advantage of this parametrization is that $ M $ interpolates smoothly
between the mass scales of the two limiting sine-Gordon models, which correspond
to $ \eta =0 $ and $ 1 $, respectively. Introducing the notation
\begin{equation}
D=\frac{(1-\eta )^{1+\frac{1}{x_{\alpha }x_{\beta }}}}{\kappa
^{\frac{1}{x_{\beta }}}_{\beta }}+\frac{\eta ^{1+\frac{1}{x_{\alpha
}x_{\beta }}}}{\kappa ^{\frac{1}{x_{\alpha }}}_{\alpha }}\,,
\end{equation}
we obtain 
\begin{equation}
\mu =  \frac{M^{x_{\beta }}(1-\eta )^{\frac{1}{x_{\alpha
}}}}{D^{x_{\beta }}}\quad , \quad
\lambda = \frac{M^{x_{\alpha }}\eta ^{\frac{1}{x_{\beta }}}}{D^{x_{\alpha }}}\,,
\end{equation}
and the Hamiltonian can be made dimensionless as follows: 
\begin{eqnarray}
h_{\text{DSG}}=\frac{1}{M}H_{\text{DSG}} & = & \frac{2\pi }{l}\left( L_{0}+\bar{L}_{0}-\frac{c}{12}+\frac{(1-\eta )^{\frac{1}{x_{\alpha }}}}{2(2\pi )^{x_{\beta }-1}}\left( \frac{l}{D}\right) ^{x_{\beta }}\left( V_{\beta }(1)+V_{-\beta }(1)\right) +\right. \nonumber \\
 &  & \left. \frac{\eta ^{\frac{1}{x_{\beta }}}}{2(2\pi )^{x_{\alpha }-1}}\left( \frac{l}{D}\right) ^{x_{\alpha }}\left( V_{\alpha }(1)+V_{-\alpha }(1)\right) \right) \,,\label{dimless_ham} 
\end{eqnarray}
where we introduced the dimensionless volume parameter 
\begin{equation}
\label{dimless_volume}
l=ML\: .
\end{equation}
We denote the eigenvalue of $ h_{\text{DSG}} $ corresponding to the
state $\Psi$ and as a function of $l$ by $\epsilon_{\Psi}(l)$ i.e. 
$E_{\Psi}(L)=M\epsilon_{\Psi}(ML)$.

It is very important that $ h_{\text{DSG}} $ interpolates smoothly
between the two extremal sine-Gordon models as $ \eta $ varies from $
0 $ to $ 1 $.  Had we chosen as our units one of the soliton mass
scales $ M_{\beta } $ or $ M_{\alpha } $, the dimensionless
Hamiltonian would have been singular at the other end of the $ \eta $
interval. We shall discuss that there is a little inconvenience in the
choice (\ref{interpolating_mass_scale}) when we want to compare with
perturbation theory around one of the end points, since such a
calculation is more naturally formulated in terms of the mass scale of
the corresponding extremal sine-Gordon model rather than the
interpolating mass scale.  However, the convenience of being able to
scan the whole range of $ \eta $ with a single numerical program well
outweighs this disadvantage.

\section{Form factor perturbation theory}\label{sec:ffpt}

\subsection{Generalities}

To supplement and check TCSA (which is a numerical method) with some
analytic results, we use the so-called \emph{form factor perturbation
theory} (FFPT).  The viewpoint of considering a nonintegrable field
theory as a perturbation of an integrable one was first taken in
\cite{nonintegrable} to which we refer the reader for a detailed
exposition of the subject. Here we restrict ourselves to a brief
summary of the necessary formulae.

Consider an integrable quantum field theory with action $ \mathcal{A}_{0} $ perturbed
by a spinless local field $ \Psi (x) $:
\begin{equation}
\mathcal{A}=\mathcal{A}_{0}-g \int \ud ^{2}x\Psi (x)\,.\end{equation}
Assume that the spectrum of the unperturbed theory $ \mathcal{A}_{0} $ is known, together
with the following matrix elements of $ \Psi (x) $ (form factors):
\begin{equation}
F_{a_{1}\dots a_{n}}^{\Psi }\left( \vartheta _{1},\dots ,\: \vartheta
_{n}\right) _{0}=\,_{0}\langle 0|\Psi (0)|a_{1}\left(
\vartheta _{1}\right) \dots a_{n}\left( \vartheta _{n}\right) 
\rangle ^{\text{in}}_{0}\,,\end{equation} 
where $ \left| 0\right\rangle _{0}
$ is the vacuum state and $ \left| a_{1}\left( \vartheta _{1}\right)
\dots a_{n}\left( \vartheta _{n}\right) \right\rangle ^{\text{in}}_{0} $
denotes an asymptotic $ n $-particle in-state of particles $
a_{1},\dots ,\: a_{n} $ with rapidities $ \beta _{1},\dots ,\: \beta
_{n} $ in the unperturbed theory $ \mathcal{A}_{0} $. Using FFPT one
can calculate the spectrum of the theory $ \mathcal{A} $
perturbatively in the coupling constant $g$.  Here we list the results
obtained to first order. The vacuum energy density is shifted by an
amount
\begin{equation}
\label{vac_energy_shift}
\delta \mathcal{E}_{\text{vac}}=g \: _{0}\langle 0| \Psi | 0\rangle _{0}\: .
\end{equation}
The mass (squared) matrix $ M^{2}_{ab} $ gets a correction
\begin{equation}
\label{mass_correction}
\delta M_{a\bar{b}}^{2}=2g F_{ab}^{\Psi }\left( i\pi \, ,\, 0\right) \delta _{m_{a},m_{b}}
\end{equation}
supposing that the original mass matrix was diagonal and of the form
\begin{equation}
M_{ab}^{2}=m^{2}_{a}\delta _{ab}\: .\end{equation}
Finally, the scattering amplitude for the four particle process $ a+b\, \rightarrow \, c+d $
is modified by 
\begin{equation}
\label{smatr_corr}
\delta S^{cd}_{ab}\left( \vartheta \right) =-ig \frac{F_{\bar{c}\bar{d}ab}^{\Psi }\left( i\pi ,\, \vartheta +i\pi ,\, 0,\, \vartheta \right) }{m_{a}m_{b}\sinh \vartheta }\quad ,\quad \vartheta =\vartheta _{a}-\vartheta _{b}\: .
\end{equation}

\subsection{Analytic results}

Using the general formulas above and the ones for the form factors and vacuum
expectation values, listed in Appendix \ref{ffs_and_vevs}, one can derive analytic
formulas for the first order corrections to the vacuum energy
density and the spectrum.

We identify the unperturbed theory and the perturbing operator in the following
way:
\begin{equation}
\mathcal{A}_{0}=\int \ud t\int \ud x\left( \frac{1}{2}\partial _{\mu }\Phi \partial ^{\mu }\Phi +\mu \, :\, \cos \beta \Phi \, :\right) \end{equation}
\begin{equation}
\Psi (x)=:\, \cos \left( \alpha \Phi +\delta \right)\, : \quad ,\quad
g=-\lambda \,.\end{equation}
We recall that 
\begin{equation}
\frac{\alpha }{\beta }=\frac{m}{n}\quad ,\quad m,\, n\in {\mathbb N}\end{equation}
and we denote the soliton mass in the unperturbed theory $ \mathcal{A}_{0} $ by
$ M_{\beta } $, as in Subsection
\ref{subsec:definition}. Now it is straightforward to apply the
general formulae summarized above, however, some care must be taken
since in the unperturbed theory ($\lambda=0$) the vacuum and the
breather states come in $n$-fold degenerate multiplets reflecting the
folded nature of the limiting SG model. When the perturbation is
switched on ($\lambda\neq 0$) it not only shifts the vacuum energy
density and the masses but also removes (possibly part of) the
degeneracies. 

Note that it is possible to interchange the role of the two cosine
terms and therefore there are eventually two perturbative regimes of
the model. We present the formulas for the above assignment; it is a
trivial matter (by suitably shifting the field and redefining the $
\delta $ phase parameter) to get the formulae corresponding to the
other possibility. We also frequently use the parameter

\begin{equation}
p=\frac{\beta ^{2}}{8\pi -\beta ^{2}}\: .\end{equation}

\subsubsection{Vacuum energy densities}

First we discuss the shift in the vacuum energy density. From (\ref{angular_nfold})
we see that the model $ \mathcal{A}_{0} $ has $ n $ vacua $ \left| k\right\rangle \: ,\: k=0,\dots ,n-1 $,
characterized by
\begin{equation}
\left\langle k\right| \Phi (x)\left| k\right\rangle =\frac{2\pi
}{\beta }k\, .\end{equation}
In infinite volume, no local operator has nonzero matrix elements
between  $ \left| k\right\rangle$ and $ \left| k'\right\rangle$ for
$k\neq k'$; therefore, the perturbative formulas can be applied for
each state $ \left| k\right\rangle$ straightforwardly. This applies to
all excited states above the vacua $ \left| k\right\rangle$ as well
(see e.g. the breathers in the next subsection).
According to (\ref{vac_energy_shift}) the vacuum energy density of the $ k $th
vacuum is shifted by 
\begin{equation}
\delta \mathcal{E}_{k}=-\frac{\lambda }{2}\left(
e^{i\delta}\mathcal{G}^{(k)}_{\alpha }(\beta
)+e^{-i\delta}\mathcal{G}^{(k)}_{-\alpha }(\beta )\right) \quad ,
\quad \mathcal{G}^{(k)}_{\alpha }(\beta)=\Big\langle k\Big|
e^{i\alpha\Phi(0)}\Big| k\Big\rangle\:
.\end{equation} Using (\ref{lz_vev}, \ref{nth_vev}) we obtain the final
form
\begin{equation}
\label{first_order_vacuum_corr}
\delta \mathcal{E}_{k}=-\lambda \mathcal{G}_{\alpha }(\beta )\cos \left( \frac{2\pi \alpha }{\beta }k+\delta \right) \: ,
\end{equation}
where $\mathcal{G}_{\alpha }(\beta)=\mathcal{G}^{(0)}_{\alpha }(\beta)$. 
We remark that one can extract a similar result from the classical potential
\begin{equation}
V(\Phi )=-\mu \cos \beta \Phi -\lambda \cos \left( \alpha \Phi +\delta
\right)\,. 
\end{equation}
At $ \lambda =0 $, the minima are
\begin{equation}
\Phi ^{(0)}_{k}=\frac{2\pi k}{\beta }\quad ,\quad V\left( \Phi ^{(0)}_{k}\right) =-\mu \: .\end{equation}
Switching on a small $ \lambda  $, the minima are shifted to 
\begin{equation}
\Phi _{k}=\Phi _{k}^{(0)}+\delta \Phi _{k}\quad ,\quad \delta \Phi _{k}=-\lambda \frac{\alpha }{\mu \beta ^{2}}\sin \left( 2\pi k\frac{\alpha }{\beta }+\delta \right) +O\left( \lambda ^{2}\right) \end{equation}
and the new values of the potential at these minima are 
\begin{equation}
V\left( \Phi _{k}\right) =-\mu -\lambda \cos \left( \frac{2\pi \alpha }{\beta }k+\delta \right) \: .\end{equation}
Therefore we obtain
\begin{equation}
\delta \mathcal{E}^{\text{classical}}_{k}=-\lambda \cos \left( \frac{2\pi \alpha }{\beta }k+\delta \right) \: ,\end{equation}
which is exactly the classical limit of (\ref{first_order_vacuum_corr}) since
it can be easily proven from (\ref{lz_vev}) that 
\begin{equation}
\mathcal{G}_{\alpha }(\beta )\, \rightarrow \, 1\quad \text{as}\quad \beta \, \rightarrow \, 0\quad ,\quad \frac{\alpha }{\beta }=\text{fixed}\, .\end{equation}

\subsubsection{Mass corrections}

We  calculate the first order corrections to the masses of the first and the
second breathers, $ M_{1} $ and $ M_{2} $. In general
\begin{equation}
\delta M_{r}=\frac{\delta M^{2}_{r}}{2M_{r}}=\frac{\delta
M_{r}^{2}}{4M_\beta\sin \frac{rp\pi }{2}}\,,
\end{equation}
where, as before, $ M_\beta $ is the soliton mass in the unperturbed
theory $\lambda=0$.

Eventually, there are $ n $ copies of each breather: $ B_{r}^{(k)} $ is
the $ r $th breather living over the vacuum $ \left| k\right\rangle \: ,\: k=0,\dots ,n-1 $
(c.f. \cite{kfold}). In the theory $ \mathcal{A}_{0} $, their masses are
independent of $ k $, but the degeneracy is split by the perturbing
potential.
We  denote the corresponding mass corrections by $ \delta M^{(k)}_{r} $.
The masses of the first and second breathers can be calculated using (\ref{mass_correction})
and (\ref{ff_11}, \ref{ff_22}) and turn out to be
\begin{eqnarray}
\delta M^{(k)}_{1} & = & \frac{\lambda \mathcal{G}_{\alpha }(\beta )\mathcal{N}}{M_{\beta }}\cos \left( \frac{2\pi \alpha }{\beta }k+\delta \right) \frac{\sin ^{2}\left( p\pi \frac{\alpha }{\beta }\right) }{\sin ^{2}\left( \frac{p\pi }{2}\right) }\exp \left\{ -\frac{1}{\pi }\int ^{p\pi }_{0}\ud t\frac{t}{\sin t}\right\}\,, \label{first_breather_corr} \\
\delta M^{(k)}_{2} & = & \frac{\lambda \mathcal{G}_{\alpha }(\beta )\mathcal{N}^{2}}{M_{\beta }}\cos \left( \frac{2\pi \alpha }{\beta }k+\delta \right) \frac{\sin ^{2}\left( p\pi \frac{\alpha }{\beta }\right) }{\sin ^{2}\left( \frac{p\pi }{2}\right) \cos \left( p\pi \right) }\times \nonumber \\
 &  & \left( 2\cos ^{2}\left( \frac{p\pi }{2}\right) -\sin ^{2}\left(
p\pi \frac{\alpha }{\beta }\right) \right) \exp \left\{ -\frac{2}{\pi
}\int ^{p\pi }_{0}\ud t\frac{t}{\sin t}\right\}\,, 
\label{second_breather_corr} 
\end{eqnarray}
(the constant $\mathcal{N}$ is defined in (\ref{Nfactor})).

One can also do a semiclassical analysis for the first breather mass. We remark
that the first breather is identified with the particle created by the fundamental
scalar field $ \Phi  $ and therefore its mass is given by the second derivative
of the potential $ V $ at its minima. Explicitly we obtain 
\begin{equation}
M^{(k)}_{1}=\sqrt{\mu }\beta +\lambda \frac{\alpha ^{2}}{2\sqrt{\mu
}\beta }\cos \left( \frac{2\pi \alpha }{\beta }k+\delta \right)
+O\left( \lambda ^{2}\right)\,, 
\end{equation}
which can be matched with the classical limit $ \beta \, \rightarrow \, 0 $
of (\ref{first_breather_corr}) using
\begin{equation}
M^{\text{classical}}_{\beta }=\frac{8\sqrt{\mu }}{\beta}\,,\end{equation}
and $\mathcal{N}\, \rightarrow \, 1$ as $ \beta \, \rightarrow \, 0$
following from (\ref{Nfactor}).

\subsubsection{S-matrix corrections}

We can also compute the first order correction to the $ B_{1}-B_{1} $ $ S $-matrix,
using (\ref{smatr_corr}, \ref{ff_1111}). The result is 
\begin{eqnarray}
\delta S_{11}(\vartheta ) & = & i\frac{\lambda \mathcal{G}_{\alpha }(\beta )\mathcal{N}^{2}}{M^{2}_{\beta }}\cos \left( \frac{2\pi \alpha }{\beta }k+\delta \right) S_{11}(\vartheta )\frac{4\cos ^{2}\left( \frac{p\pi }{2}\right) \sin ^{2}\left( p\pi \frac{\alpha }{\beta }\right) }{\sin ^{2}\left( \frac{p\pi }{2}\right) }\times \\
 &  & \left( \frac{\sin ^{2}\left( p\pi \frac{\alpha }{\beta }\right) }{\sin ^{2}\left( p\pi \right) }-\frac{1}{\cosh (\vartheta )+1}\right) \frac{\sinh \vartheta }{\sinh ^{2}\vartheta +\sin ^{2}\left( p\pi \right) }
\end{eqnarray}
where 
\begin{equation}
S_{11}(\vartheta )=\frac{\sinh \vartheta +i\sin p\pi }{\sinh \vartheta -i\sin p\pi }\end{equation}
is the sine-Gordon $ B_{1}-B_{1} $ S-matrix. It turns out, however,
that this correction is too small and cannot be measured within the
precision of TCSA.

\section{Comparison of FFPT with TCSA}
\label{sec:comparison}

\subsection{Extracting the vacuum energy densities and the mass spectrum from TCSA}

TCSA gives the spectrum of the dimensionless Hamiltonian $ h_{TCSA} $ (\ref{dimless_ham})
as a function of the dimensionless volume parameter $ l $ (\ref{dimless_volume}).
In our numerical calculations, we use the interpolating mass scale $ M $
(\ref{interpolating_mass_scale}) as a unit of mass. However, the FFPT results
are expressed in terms of the scale $ M_{\beta } $; we must therefore take
care of converting between the two conventions. We describe first how
we extract the vacuum energy densities and the various masses from the
TCSA data then we compare these \lq experimentally measured'
quantities and the FFPT predictions on a specific example when $\alpha
/\beta =1/2$.

\subsubsection{Vacuum energy density}

Concerning the vacuum energy density, our numerical results show that the ground
state energy is linear in the volume $ L $
\begin{equation}
E_{\text{vac}}=\mathcal{E}_{\text{vac}}L\end{equation}
for a wide range of $ L $. Therefore the (dimensionless) vacuum energy density
$\mathcal{E}_{\text{vac}}/M^{2}$ can be 
measured as $\epsilon _{\text{vac}}\left( l_{0}\right)/l_0$
by choosing some value $ l_{0} $ in this range (typically between $ 14 $
and $ 20 $). We extract the shift in the vacuum energy density using the relation
\begin{equation}
\frac{\delta \mathcal{E}_{\text{vac}}}{M^{2}_{\beta }}=\frac{\epsilon _{\text{vac}}\left( l_{0}\right) }{l_{0}}\left( \frac{M}{M_{\beta }}\right) ^{2}-\frac{\mathcal{E}^{(0)}_{\text{vac}}}{M^{2}_{\beta }}\end{equation}
where $ \mathcal{E}^{(0)}_{\text{vac}} $ is the value measured using TCSA in the
unperturbed theory at $ \eta =0 $. It's value is known analytically as well
\begin{equation}
\frac{\mathcal{E}^{(0)}_{\text{vac}}}{M^{2}_{\beta }}=-\frac{1}{4}\tan \frac{p\pi }{2}\: ,\end{equation}
however, we prefer to use the value extracted numerically because it helps eliminating
part of the truncation error coming from TCSA.

\subsubsection{Masses}

We know that the energy of a particle $ a $ of momentum $ 0 $ in finite
volume, relative to the ground state,  has the form
\begin{equation}
E_{a}(L)-E_{vac}(L)=M_{a}+O\left( e^{-ML}\right) 
\label{mass_as}
\end{equation} 
where $ M_{a} $ is the mass of the particle and $ M $ is some mass scale
characterizing the theory. As the interpolating scale
(\ref{interpolating_mass_scale}) agrees with the relevant scales at
the two extremal points $ \eta =0 $ and $ \eta =1 $, while in between
it interpolates smoothly between them, we can take it as the
characteristic scale of the DSG model. Therefore the simplest way to
measure the mass is to extract it as the value of the energy above the
ground state at a suitable value of the volume in a range (the scaling
region) where the energy difference $E_{a}(L)-E_{vac}(L)$ is
approximately constant; (however, this volume cannot be too large since
then truncation errors would spoil the measurement). Therefore to obtain the
mass corrections we use the relation
\begin{equation}
\frac{\delta M_{a}}{M_{\beta }}=\left( \epsilon _{a}\left( l_{0}\right) -\epsilon _{\text{vac}}\left( l_{0}\right) \right) \frac{M}{M_{\beta }}-\frac{M^{(0)}_{a}}{M_{\beta }}\end{equation}
where $l_{0}$ is a suitable value of the volume parameter, $\epsilon _{a}(l_{0})$
is the relevant eigenvalue of (\ref{dimless_ham}) and $M^{(0)}_{a}$ is
the mass measured at $ \eta =0 $. Once again, the mass ratios 
\begin{equation}
\frac{M^{(0)}_{a}}{M_{\beta }}\end{equation}
 are known analytically, but we use the values extracted from TCSA in order
to eliminate part of the numerical errors.

\subsection{A special case: $ \alpha=\beta /2 $}\label{subsec:DSG2}

We  present the comparison between the numerical TCSA data and the analytical
first order FFPT results for a special ratio of the frequencies. The reason
for choosing this example is that (1) it is the simplest possible case and (2)
this is the one that we treat nonperturbatively in detail later on. Let us first
discuss the model itself. The action is
\begin{equation}
\mathcal{A}=\int \ud t\int \ud x\left( \frac{1}{2}\partial _{\mu }\Phi
\partial ^{\mu }\Phi +\mu \cos \beta \Phi +\lambda \cos \left(
\frac{\beta }{2}\Phi +\delta \right) \right) 
\end{equation} 
We denote
this model by $ \text{DSG}^{\eta }_{2}(\beta ,\delta ) $ (sometimes
omitting the parameters in the parentheses). The field is defined as
an angular variable (\ref{angular_nfold}), in this case with the
period
\begin{equation}
\Phi \equiv \Phi +\frac{4\pi }{\beta }l\quad ,\quad l\in {\mathbb Z}\: .\end{equation}
Therefore the theory $ \text{DSG}^{\eta =0}_{2} $ has two vacua, which
are degenerate in infinite volume and correspond to the classical minima
\begin{equation}
\Phi _{0}=0\quad ,\quad \Phi _{1}=\frac{2\pi }{\beta }\ ,\end{equation}
thus this model corresponds to a $ 2 $-folded sine-Gordon theory $ \text{SG}(\beta ,2) $
\cite{kfold}. The other extremal case $ \text{DSG}^{\eta =1}_{2} $
has only one vacuum
\begin{equation}
\Phi _{0}'=-\frac{2\delta }{\beta }\end{equation}
 and is a $ 1 $-folded model $ \text{SG}(\beta/2 ,1) $.

In the vacuum sector of $ \text{SG}(\beta ,2) $, every state essentially
appears in two copies, corresponding to the two different vacua. In infinite
volume, these copies are degenerate, but the degeneracy is lifted for finite
$ L $ by the tunnelling between the two vacua. In particular, the vacua and
the (neutral) one-particle states are split by an amount decreasing exponentially
with the volume $ L $. Switching on the perturbation removes this
degeneracy in an interesting way.

Let us first discuss the case $ \delta =0 $. Switching on $ \lambda $,
the two vacua have different vacuum energy densities and thus the
leading term in the split between the two ground states becomes a term
linear in $ L $. To first order in $ \lambda $, we obtain from
(\ref{first_order_vacuum_corr})
\begin{equation}
E^{(1)}_{\text{vac}}(L)-E^{(0)}_{\text{vac}}(L)=2\lambda
\mathcal{G}_{\beta /2}(\beta )L+\dots 
\end{equation}
where $ E^{(0,1)}(L) $ are the energies of the two vacuum states
(Casimir energies) as functions of $ L $ and the ellipses denote terms
which fall off exponentially with the volume. Since the vacuum energy
density is universal for all excitations lying above the same ground
state, the space of states separates into two parts which manifest
themselves as lines with two different slopes in the large volume
limit. In infinite volume, the Hilbert space of the system eventually
reduces to the ones with the smaller slope, as the other set of states
will lie at infinitely high energy above them. For this reason we call the
states with the larger value of the vacuum energy density
\emph{``runaway''} states.

\begin{figure}
\begin{center}
 \subfigure[$\eta=0$]{
  \resizebox*{0.4\textwidth}{0.2\textheight}{\includegraphics{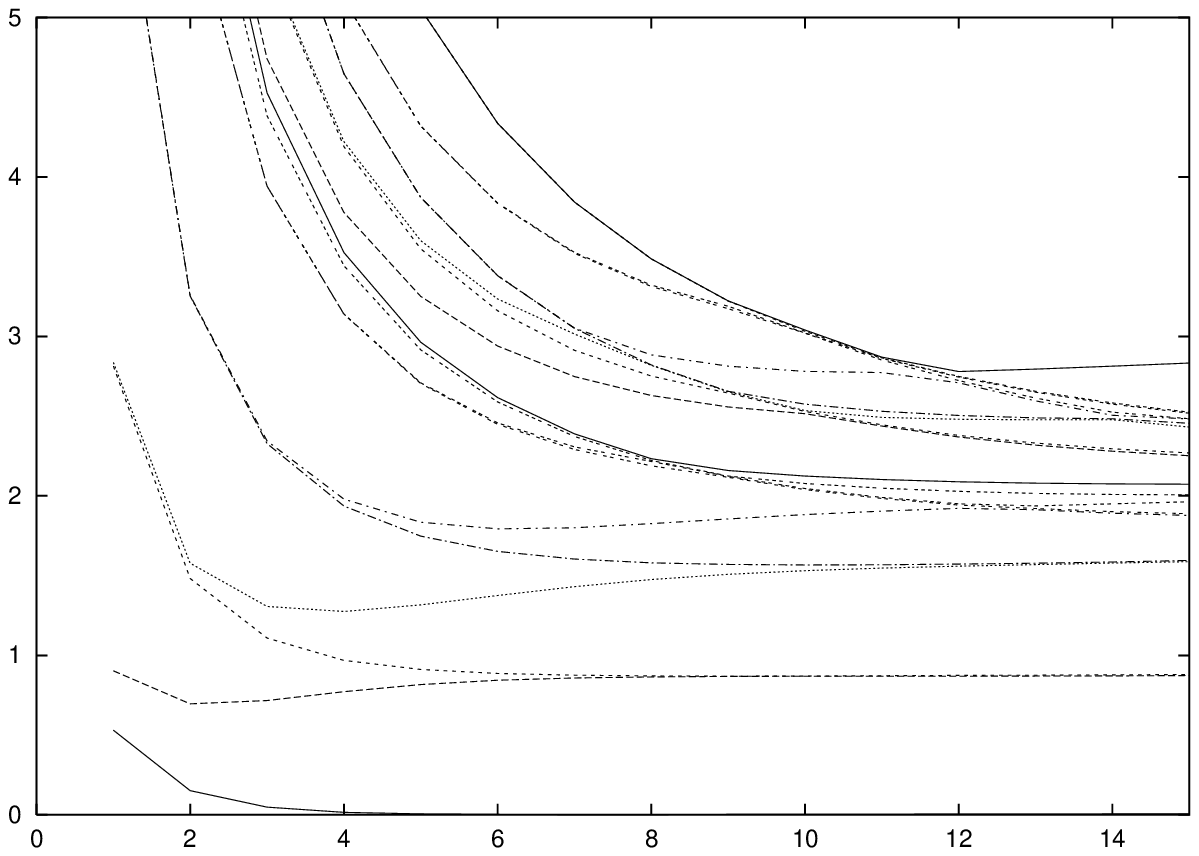}}
 }
 \subfigure[$\eta=0.2$]{
  \resizebox*{0.4\textwidth}{0.2\textheight}{\includegraphics{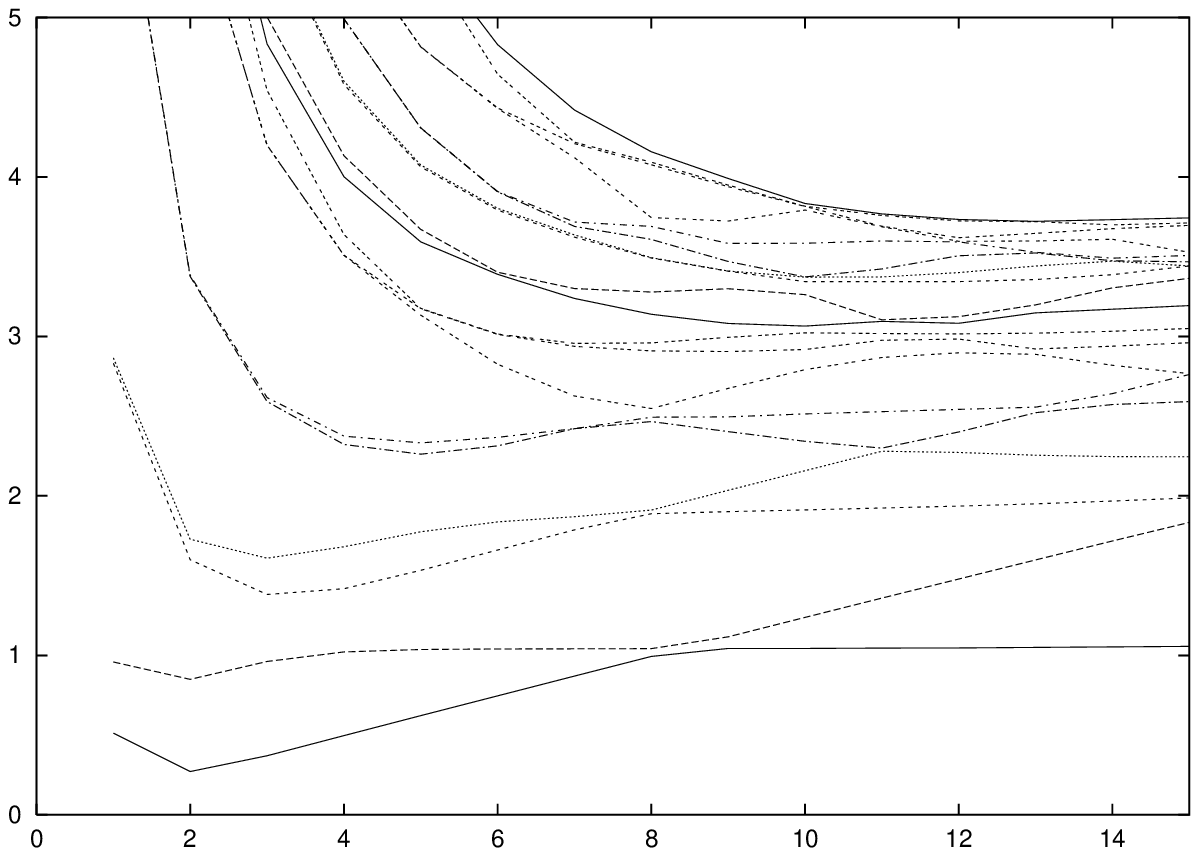}}
 }\\
 \subfigure[$\eta=0.4$]{
  \resizebox*{0.4\textwidth}{0.2\textheight}{\includegraphics{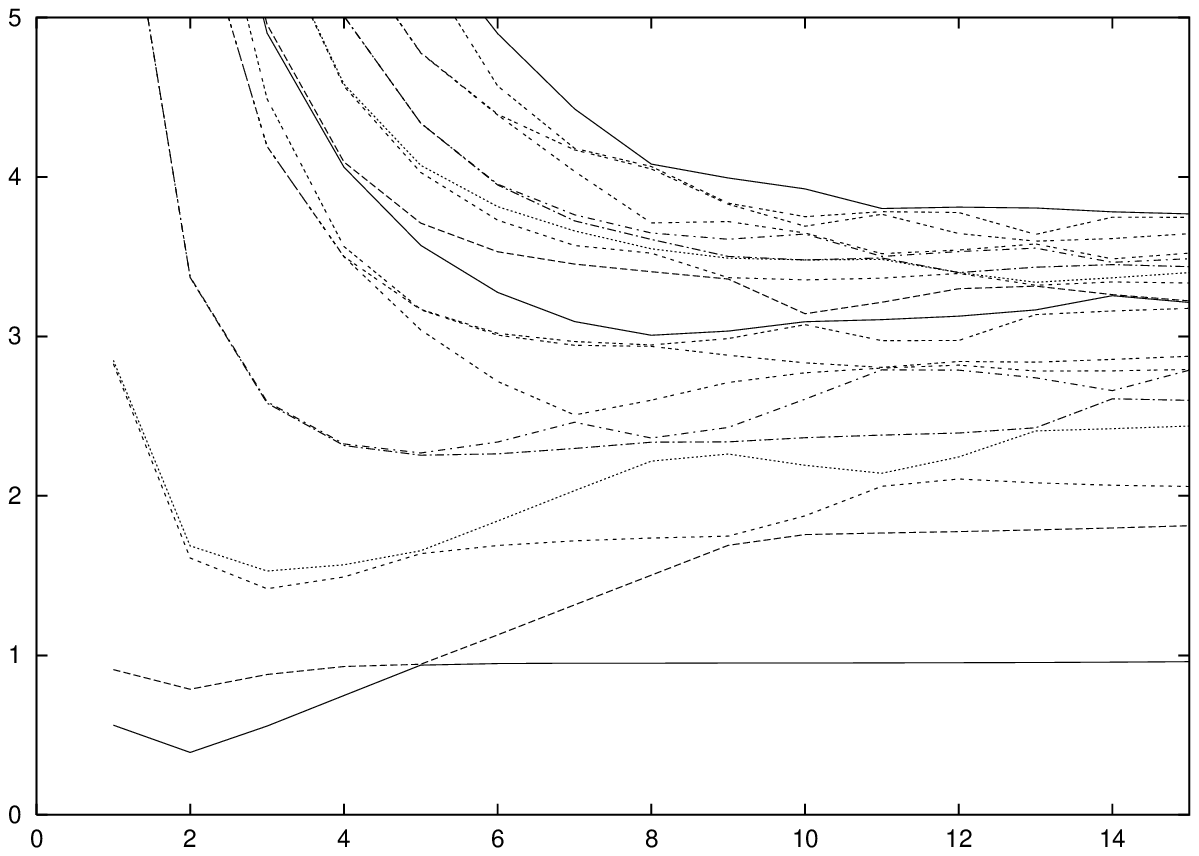}}
 }
 \subfigure[$\eta=0.6$]{
  \resizebox*{0.4\textwidth}{0.2\textheight}{\includegraphics{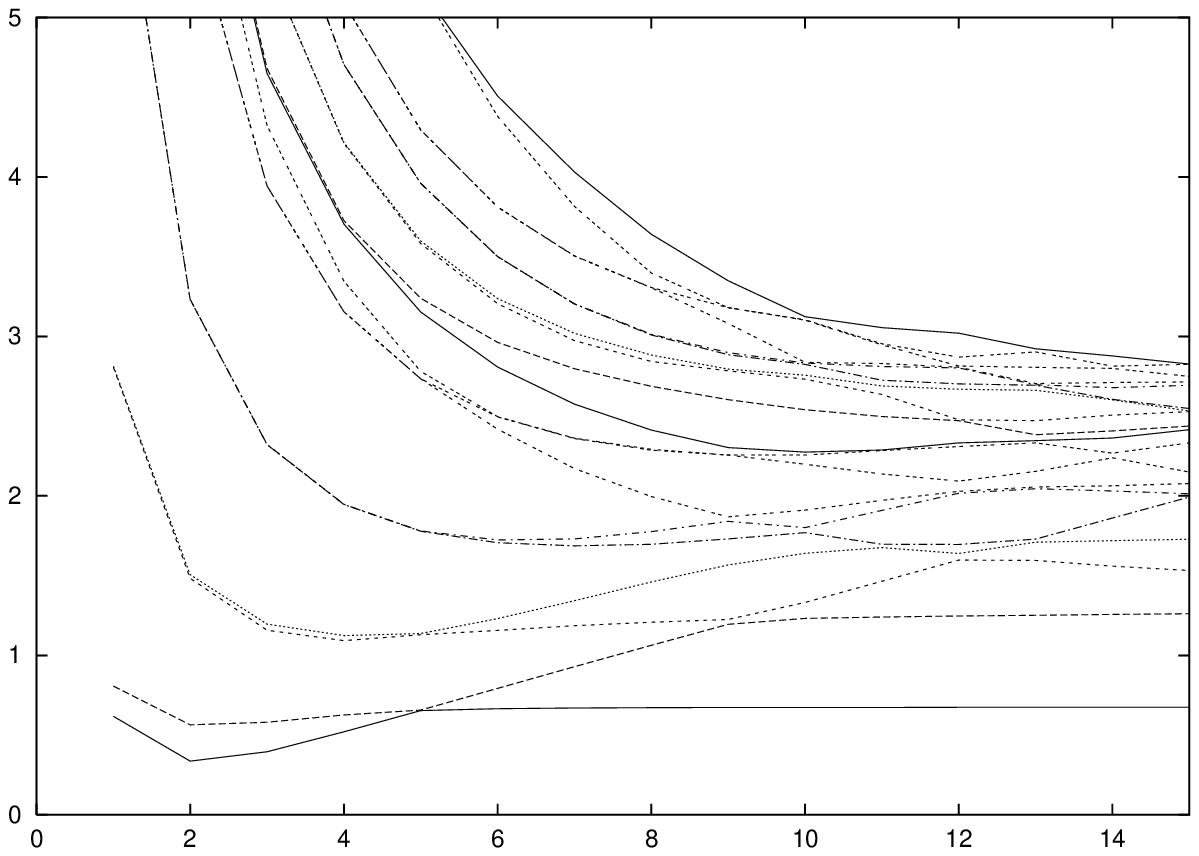}}
 }
 \subfigure[$\eta=0.8$]{
  \resizebox*{0.4\textwidth}{0.2\textheight}{\includegraphics{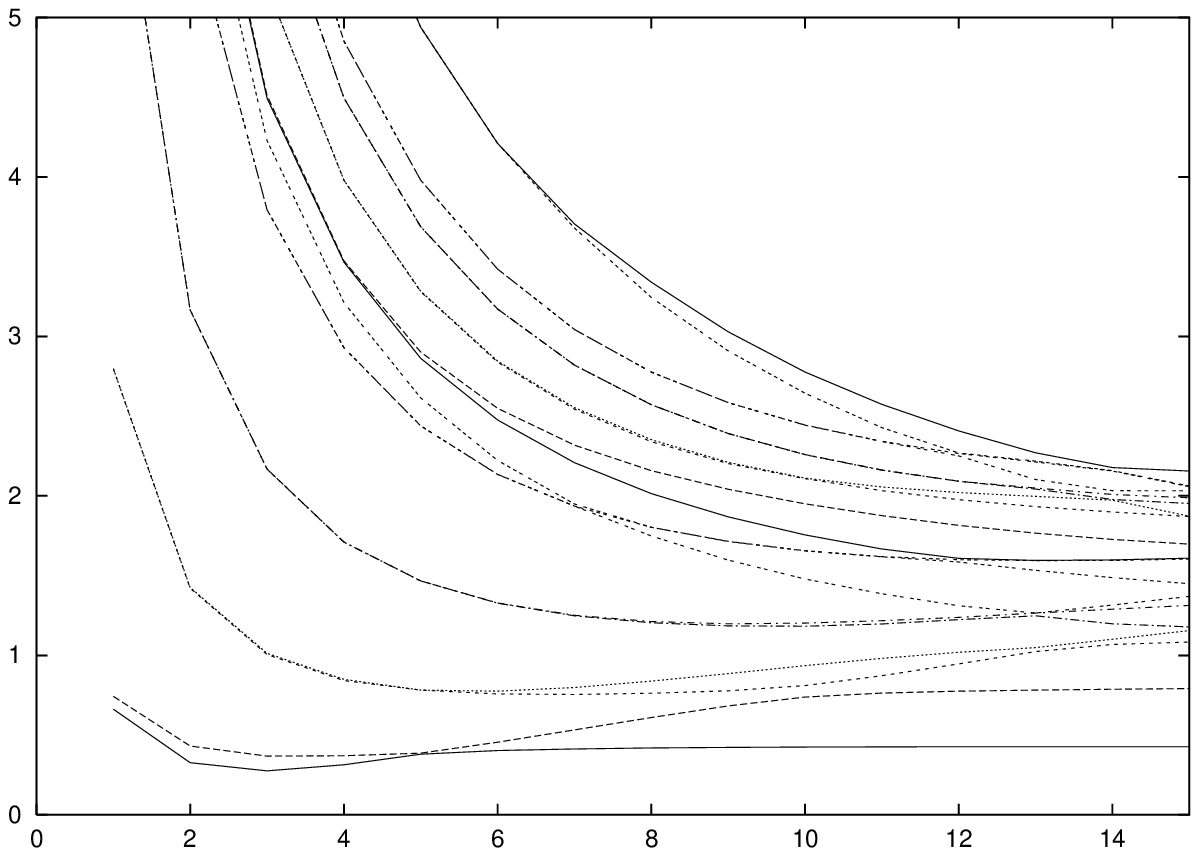}}
 }
 \subfigure[$\eta=1$]{
  \resizebox*{0.4\textwidth}{0.2\textheight}{\includegraphics{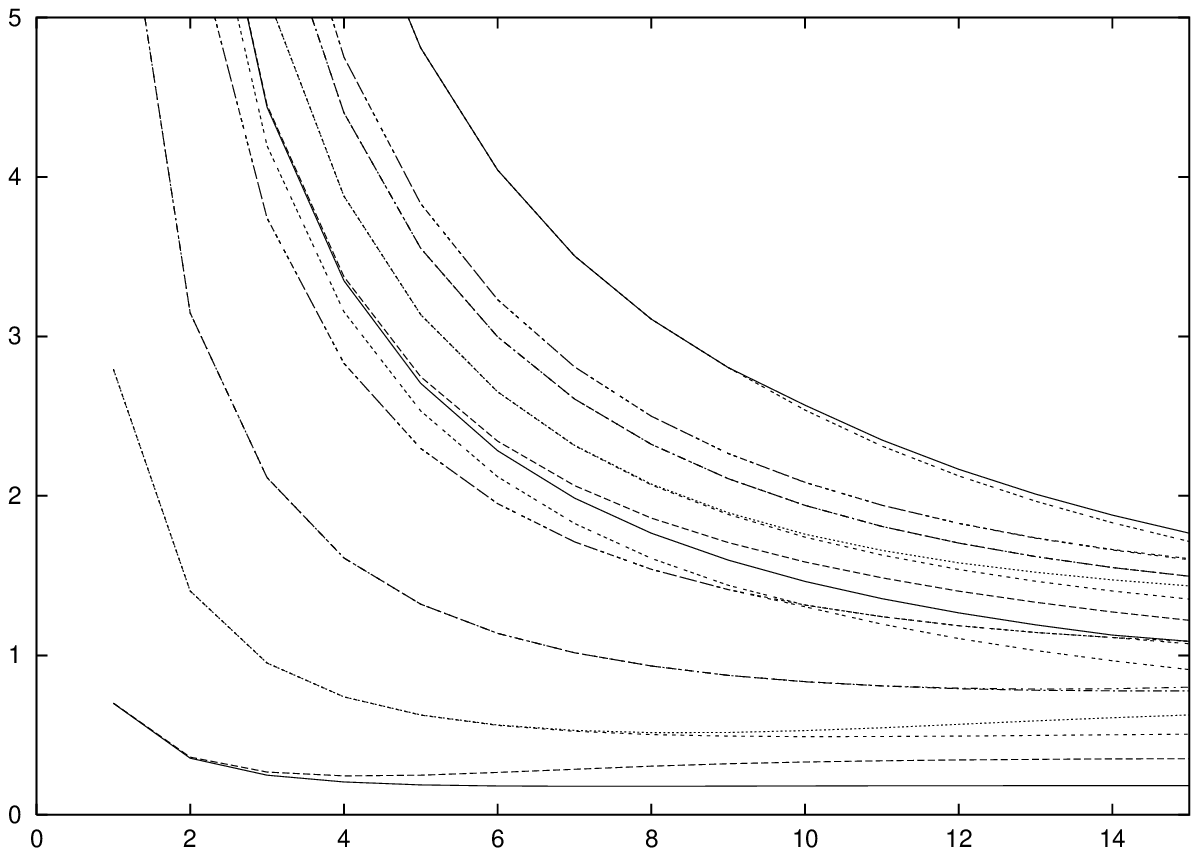}}
 }
\end{center}
\caption{Change of the spectrum as $\eta$ varies from $0$ to $1$ in
$\text{DSG}_2^{\eta}$ for $\beta=4\sqrt{\pi}/3$ and $\delta=0$. The
energies are normalized by subtracting the ground state
contribution. The plots show the first $20$ states (including the
ground state, which has constant zero energy in this convention).}
\label{fig:movie}
\end{figure}

Of course, higher order terms modify the linear dependence of the
slope on $\lambda$.  In the other extremal point $ \text{DSG}^{\eta
=1}_{2} $ one has only one vacuum, but, as the corresponding
sine-Gordon model is more attractive, it has more breather
states. This is where the interpolating mass parameter
(\ref{interpolating_mass_scale}) comes in handy: as the Hamiltonian
(\ref{dimless_ham}) depends smoothly on $ \eta $, one can run a \lq
movie' in $ \eta $, diagonalizing the Hamiltonian in TCSA for
different values of $ \eta $ from $ 0 $ to $ 1 $. As shown in Figure
\ref{fig:movie}, the energy of the set of runaway states first moves
upwards, but then turns back and comes down to form the spectrum of
the endpoint $ \text{DSG}^{\eta =1}_{2}=\text{SG}(\beta /2,1) $. 

We wish to note that the runaway states do not eventually cross the
other lines; rather, there are line avoidances which is a typical
pattern for states that are metastable in finite volume and eventually
decay \cite{luscher} (see also \cite{nonintegrable} for similar
results for the off-critical Ising model in magnetic field).

Let us now discuss what happens for general $ \delta  $. Specializing (\ref{first_order_vacuum_corr},
\ref{first_breather_corr}, \ref{second_breather_corr}), we obtain
\begin{eqnarray}
\delta \mathcal{E}^{(0)} & = & -\delta \mathcal{E}^{(1)}=-\lambda \mathcal{G}_{\beta /2}(\beta )\cos \delta \label{DSG2_vac_corr} \\
\delta M_{1}^{(0)} & = & -\delta M_{1}^{(1)}=\frac{\lambda \mathcal{G}_{\beta /2}(\beta )}{M_{\beta }}\mathcal{N}\cos \delta \exp \left\{ -\frac{1}{\pi }\int ^{p\pi }_{0}\ud t\frac{t}{\sin t}\right\} \label{DSG2_B1_corr} \\
\delta M_{2}^{(0)} & = & -\delta M_{2}^{(1)}=\frac{\lambda \mathcal{G}_{\beta /2}(\beta )}{M_{\beta }}\mathcal{N}^{2}\cos \delta \frac{3\cos ^{2}\left( \frac{p\pi }{2}\right) -1}{\cos p\pi }\exp \left\{ -\frac{2}{\pi }\int ^{p\pi }_{0}\ud t\frac{t}{\sin t}\right\} \label{DSG2_B2_corr} 
\end{eqnarray}
Note that at $ \delta =\frac{\pi }{2} $ the $ O(\lambda ) $ corrections
vanish and in particular the two vacua become degenerate. We shall show shortly
that this is also true nonperturbatively, at least as long as $ \eta  $ is smaller
than some critical value $ \eta _{\text{crit}} $.

\begin{figure}[tbp]
\begin{center}
\input{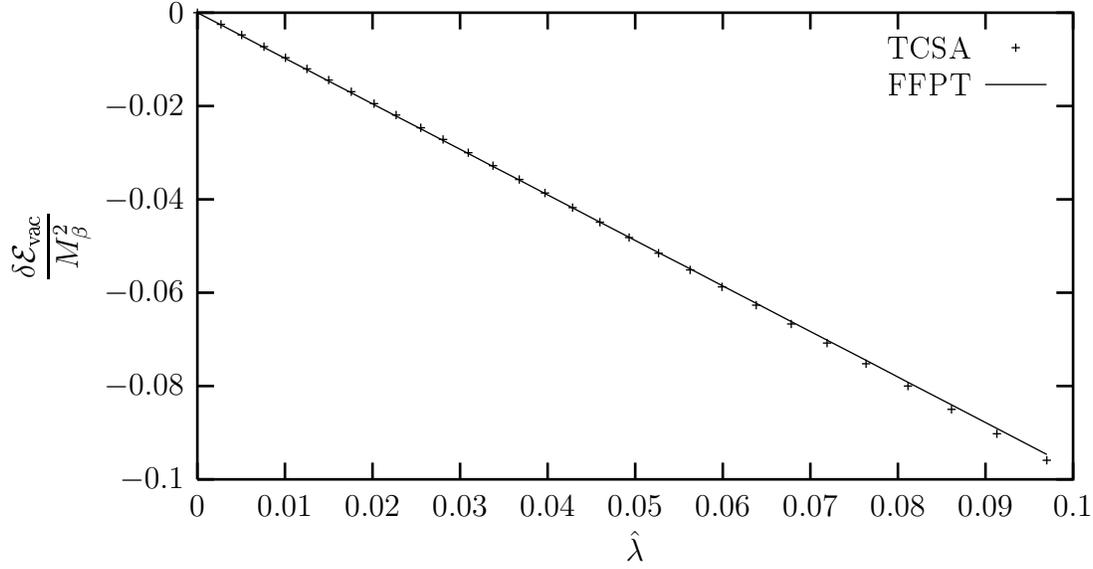}
\end{center}
\caption{The correction to the vacuum energy density as a function of $ \hat{\lambda }$}
\label{fig:vac_split}
\end{figure}

\begin{figure}[tbp]
\begin{center}
\input{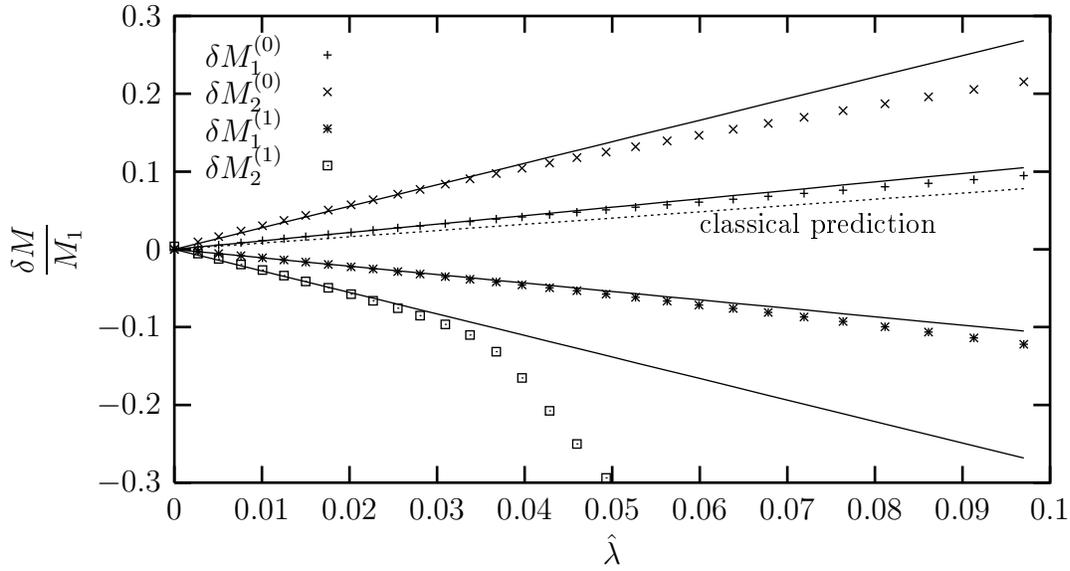}
\end{center}
\caption{Breather mass corrections as functions of $ \hat{\lambda }$}
\label{fig:br_mass_corr}
\end{figure}

\begin{figure}[tbp]
\begin{center}
\input{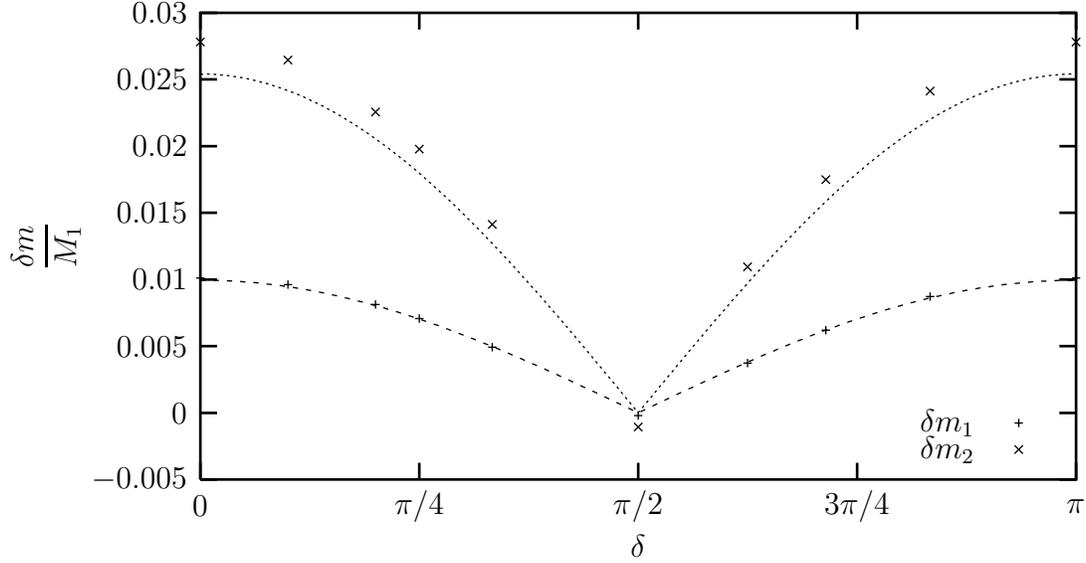}
\end{center}
\caption{Breather mass corrections as functions of 
$\delta$ 
at $\hat{\lambda}\approx 0.01$ and $\beta=4\sqrt{\pi}/3$}
\label{fig:delta_dep}
\end{figure}

Now we proceed to the comparison between FFPT and TCSA, using plots. The
first order corrections are linear in the coupling constant of the
perturbing operator, so to use a convenient and dimensionless
parametrization, we  define the following combination:
\begin{equation}
\hat{\lambda}=\lambda M_{\beta}^{-x_\beta}\ .
\end{equation}
Figure \ref{fig:vac_split} shows the results for the correction to the vacuum energy
density $\delta\mathcal{E}_{\rm vac}/M_{\beta}^2$ for
$\delta=0$ and $\beta=4\sqrt{\pi}/3$, in units of the soliton mass $M_\beta$. The agreement is
spectacular for a quite long range of the coupling.
Similarly, Figure \ref{fig:br_mass_corr} shows the breather mass
corrections for the four particles $B^{(0)}_1$, $B^{(0)}_2$,
$B^{(1)}_1$ and $B^{(1)}_2$ at the same values of $\beta$ and
$\delta$. 
For comparison we have also plotted the
classical prediction for $\delta M_1^{(0)}$.
Finally, Figure \ref{fig:delta_dep} presents the dependence of the
mass corrections on the phase parameter $\delta$ at a suitably small
value of $\hat{\lambda}$. (Note that here, to ensure better
visibility, we changed the scale on the vertical axis with respect to 
Figure \ref{fig:br_mass_corr}. As a result, the deviation between the
first order FFPT and the TCSA results is much more pronounced).

\begin{figure}[tbp]
\begin{center}
{\resizebox*{0.8\textwidth}{0.3\textheight}{\rotatebox{270}{\includegraphics{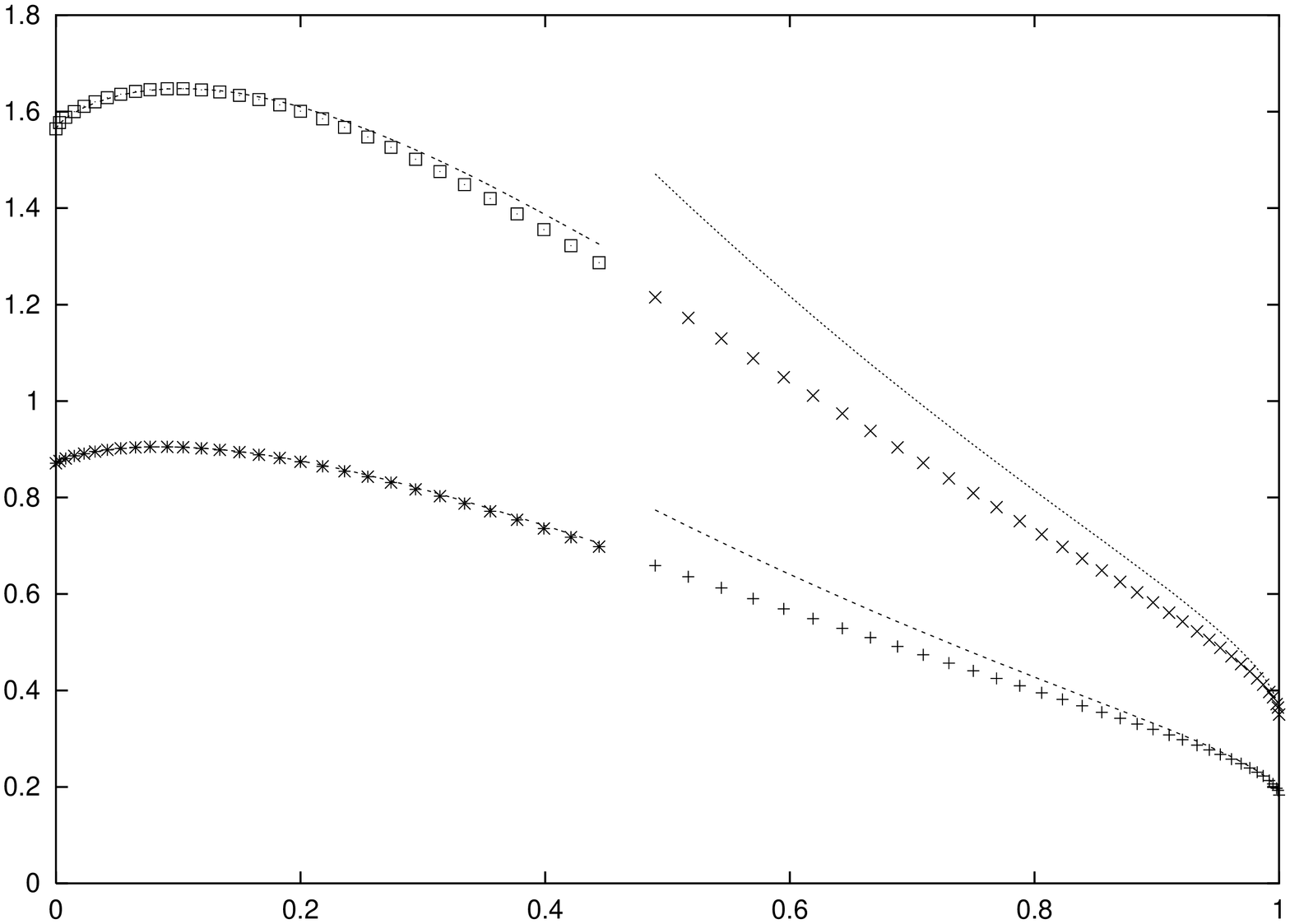}}}}
\end{center}
\caption{The dependence of the first and second breather masses on
$\eta$ running from $0$ to $1$ in
$\text{DSG}_2^\eta(4\sqrt{\pi}/3,0)$. The dots are the numerical TCSA
data while the lines are the perturbative results obtained by using
FFPT around the two endpoints $\eta=0$ and $\eta=1$.}
\label{fig:eta_dep}
\end{figure}

One can also check the agreement between TCSA and FFPT for a perturbation
theory in the coupling $\mu$, i.e. around $\eta=1$. The results are
similar, except that the validity range of first order perturbation
theory is smaller than around $\eta=0$. The dependence of the first
and second breather masses on $\eta$ is shown in Figure 
\ref{fig:eta_dep}.

To sum up, the comparison between TCSA and FFPT gives a convincing
evidence for the validity of both approaches in the perturbative
($\eta\sim 0$ or $\eta\sim 1$) domain. This is of utmost importance
because for the purposes of further investigations we shall continue
to use TCSA inside the complete interval $0<\eta<1$, where there are
no ways to check its validity by any independent method (except some
semiclassical analysis and qualitative arguments). 
 
\section{Phase transition in $ \text{DSG}^{\eta }_{2}$: generalities}
\label{sec:phasetransitions}

We now turn to examining the phase diagram of the theory
$\text{DSG}^{\eta }_{2}$ as $\eta$ varies from $0$ to $1$.
We have seen that the theory has two degenerate vacua at $\eta=0$, while
there is only one ground state at $\eta=1$. This hints at the
possibility of a phase transition happening somewhere in
between. However, for generic values of the parameters the degeneracy
between the two vacua is lifted as soon as $\eta>0$ thus the end
point $\eta=0$ is singular in this sense.

However, as we have already pointed out (subsection
\ref{subsec:DSG2}) the first order corrections to the vacuum energy
densities (\ref{DSG2_vac_corr}) vanish exactly when $\delta$ takes a
value on the boundary of its fundamental range 
\begin{equation}
\delta=\frac{\pi}{2}\,.
\end{equation}
From now on we restrict ourselves to this special value of $\delta$ and
by a slight abuse of notation denote this model by
$\text{DSG}_2^\eta$.

In this section we present an argument that the energy difference
between the two vacua eventually vanishes to all orders in $\lambda$
and that there is a critical value $\eta=\eta_{\text{crit}}$ at which there is a
second order phase transition. Since all perturbative corrections to
the vacuum energy difference vanish, this transition is entirely in a
nonperturbative regime and form factor perturbation theory is not
applicable. We first present a semiclassical (mean
field/Landau-Ginzburg) analysis, extending the arguments given by
Delfino and Mussardo \cite{delfino_mussardo} and Fabrizio et
al. \cite{nersesyan}, and discuss the signatures of first and second
order phase transitions in finite volume. In the next section we use
TCSA to verify the predictions of the semiclassical considerations and
in particular to establish the second order nature of the phase
transition and its corresponding universality class.

\subsection{Landau-Ginzburg analysis}

\begin{figure}[tbp]
\psfrag{lambda<4mu}{$\lambda<4\mu$}
\psfrag{lambda>4mu}{$\lambda>4\mu$}
\psfrag{V(Phi)}{$V(\Phi)$}
\psfrag{Phi}{$\Phi$}
\subfigure{
\resizebox*{0.5\textwidth}{0.3\textheight}{\includegraphics{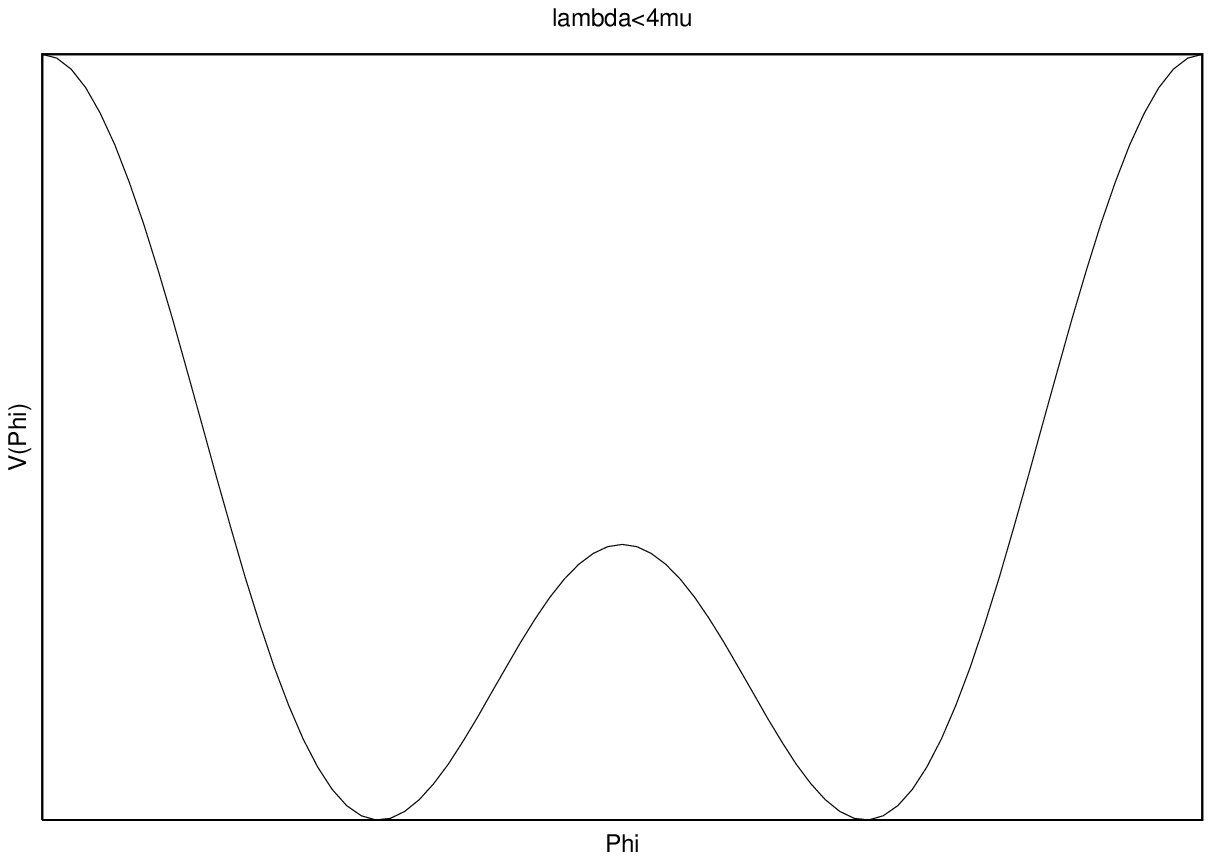}}
}
\subfigure{
\resizebox*{0.5\textwidth}{0.3\textheight}{\includegraphics{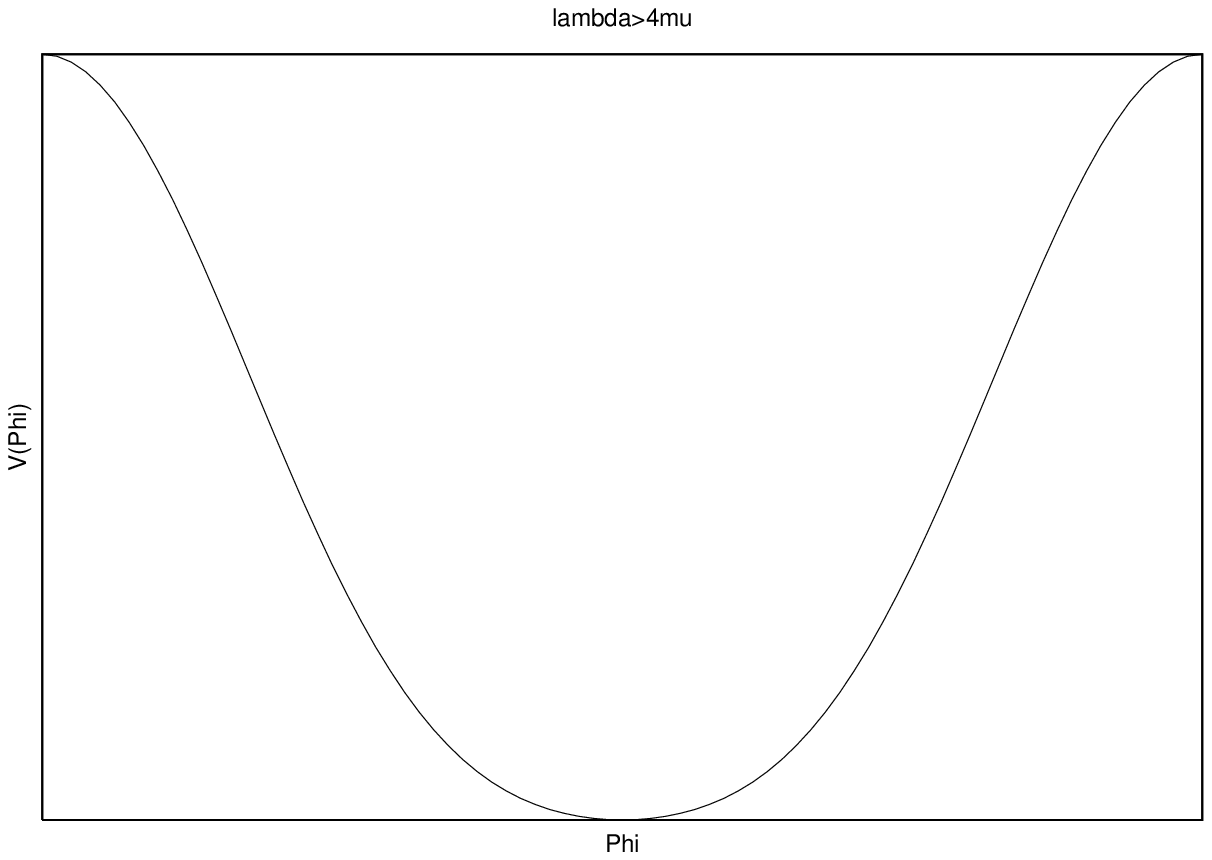}}
}
\caption{The behaviour of the classical potential $V(\Phi)$ for $\lambda<4\mu$
and  $\lambda>4\mu$}
\label{fig:pht_pot}
\end{figure}

We start our analysis with the classical potential
\begin{eqnarray}
V(\Phi )&=&-\mu \cos \beta \Phi -\lambda \cos \left( \alpha \Phi
+\delta \right) \nonumber \\
&=&-\mu \cos \beta \Phi +\lambda \sin \left( \frac{\beta}{2} \Phi\right)\,.
\end{eqnarray}
For definiteness, let us suppose that $\mu$ and $\lambda$ are positive
without losing generality.
As the potential is periodic under
$\Phi\:\rightarrow\:\Phi+4\pi/\beta$, the fundamental range of the
field variable can be taken as
\begin{equation}
-\frac{3\pi}{\beta}\leq \Phi<\frac{\pi}{\beta}\,.
\end{equation}
Note that the potential also has the discrete (${\mathbb Z}_2$)
symmetry
\begin{equation}
T:\,\Phi\:\rightarrow\:-\frac{2\pi}{\beta}-\Phi\,.\label{Z2}
\end{equation}
Then a simple analysis establishes that the potential shows the
characteristics of a second order phase transition at the point
$\lambda=4\mu$ (see Figure \ref{fig:pht_pot}). Since classically the
dimensions of $\mu$ and $\lambda$ are $x_{\alpha}=x_{\beta}=2$, the
definition (\ref{eta_definition}) of $\eta$ reduces to
\begin{equation}
\eta=\frac{\lambda ^{2}}{\mu ^{2}+\lambda ^{2}}
\end{equation}
and so this point corresponds to the following critical value of $\eta$
\begin{equation}
\eta_{\text{crit}}(\beta =0)=\frac{16}{17}=0.941\dots\ .
\label{etac_semicl}
\end{equation}

In view of the ${\mathbb Z}_2$ symmetry (\ref{Z2}) one would like to
conclude that this is a second order phase transition point in the
Ising universality class. However, there is a caveat here, as
discussed by Fabrizio et al. \cite{nersesyan}.

The essence of the problem can be summarized as follows. The potential
$V(\Phi)$ is renormalized at the quantum level. To determine the phase
structure, one needs to look at the effective potential. Due to the
periodicity and the ${\mathbb Z}_2$ symmetry, it can be readily
checked that the only possible new terms are of the form
\begin{equation}
\cos\left(r\beta\Phi\right)\,,\ r\in{\mathbb Z} \quad\text{or}\quad
\sin\left(r\beta\Phi\right)\,,\ r\in{\mathbb Z+\frac{1}{2}}\,. 
\label{higher_ops}
\end{equation}

The conformal dimensions of these operators are 
\begin{equation}
\Delta^\pm_r=\frac{r^2\beta^2}{8\pi}\ ,
\end{equation}
and so as one increases $r$ they are less and less relevant. The first
two possibilities are particularly interesting. Fabrizio et al.
\cite{nersesyan} claim that while
$\sin\left(\frac{3\beta}{2}\Phi\right)$ has no effect on the nature of
the phase transition (apart from modifying the position of the
critical point), the term $\cos\left(2\beta\Phi\right)$ which is a
relevant perturbing operator for $\beta^2<2\pi$ can (depending on its
coefficient) turn the second order phase transition into a first order
one.

Let us analyze the effect of such operators now. We take as our quantum
effective potential the expression
\begin{equation}
V_{\text{eff}}(\Phi )=-\mu \cos \beta \Phi +\lambda \sin \frac{\beta}{2}
\Phi+ \kappa\sin \frac{3\beta}{2}
\Phi+ \nu \cos 2\beta \Phi\,.
\end{equation}
$\kappa$ and $\nu$ can eventually be expressed in terms of $\lambda$
and $\mu$, but their precise expressions are unknown. 
The conditions for the existence of 
a second-order phase transition point are
\begin{equation}
V''_{\text{eff}}(\Phi)\Big|_{\Phi=-\frac{\pi}{\beta}}=0\quad
,\quad
V^{(4)}_{\text{eff}}(\Phi)\Big|_{\Phi=-\frac{\pi}{\beta}}>0\ ,
\end{equation}
since the ${\mathbb Z}_2$ symmetric extremum is at
$\Phi=-\frac{\pi}{\beta}$.

Explicitly, we have the conditions
\begin{equation}
-4\mu+\lambda-9\kappa-16\nu=0\quad , \quad
 16\mu-\lambda+81\kappa+256\nu>0\ .
\label{2ndordercond}\end{equation}
Putting $\kappa=\nu=0$ we recover that the phase transition point in
the classical potential is at $\lambda=4\mu$ and the second derivative
is positive (we assumed that $\lambda$ and $\mu$ are positive).

Suppose now that $\kappa$ and $\nu$ are independent parameters and
that we vary only $\lambda$. The critical point can be read
off from the first condition in (\ref{2ndordercond}):
\begin{equation}
\lambda_c=4\mu+9\kappa+16\nu\ .
\end{equation}
Therefore the condition for a second order transition is
\begin{equation}
\mu+6\kappa+20\nu>0\ .
\label{2ndordercond1}\end{equation}
We can draw three conclusions. First, there is ample space for
second order phase transition even in the presence of the higher
frequency terms in the effective potential. Second, the term $\sin
\frac{3\beta}{2}\Phi$ is also capable of inducing a first order
transition, since a large negative $\kappa$ is enough. Third, it is
also possible that higher order derivatives vanish at the
transition as well, giving tricritical and in fact even tetracritical points
(even higher frequency terms open the possibility to multicritical
points of any order). These observations can also be confirmed by
simply plotting the effective potential for different values of the
parameters.

However, for the double sine-Gordon model (\ref{DSG_action}) the
coefficients of the higher frequency terms (e.g. $\kappa$, $\nu$) are
fixed by the dynamics. Therefore this model lives on a specific
surface in the multiparameter space of possible quantum effective
potentials that respect the ${\mathbb Z}_2$ symmetry
$\Phi\:\rightarrow\:-\frac{2\pi}{\beta}-\Phi$ and the periodicity
$\Phi\:\rightarrow\:\Phi+4\pi/\beta$. As we have already shown, the
phase transition point is out of the reach of perturbative techniques
and therefore one needs to recourse to some nonperturbative analysis
to solve the problem. In the special case when $p=1$ ($\beta^2=4\pi$),
by using a mapping onto a generalized Ashkin Teller model, 
Fabrizio et al. \cite{nersesyan} were able to show the existence of a
second order phase transition.  
In the general case, when $0<p<1$, since the model is nonintegrable,
the only available candidate is TCSA (the upper limit is to  
avoid any UV problems with TCSA).

\begin{figure}
\psfrag{V(Phi)}{$V(\Phi)$}
\psfrag{Phi}{$\Phi$}
\begin{center}
 \subfigure[$\lambda<\frac{3}{5}\mu$]{
  \resizebox*{0.4\textwidth}{0.2\textheight}{\includegraphics{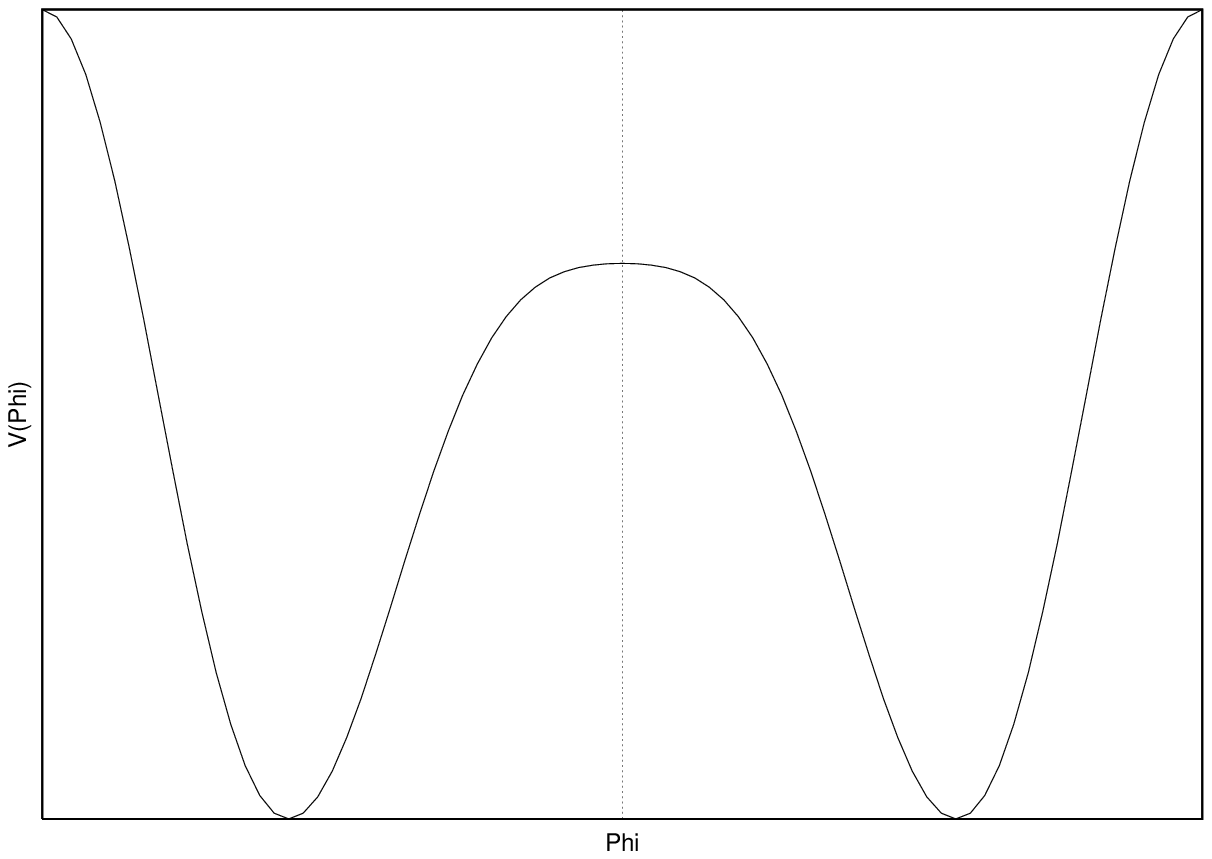}}
 }
 \subfigure[$\frac{3}{5}\mu<\lambda<\frac{13}{8}\mu$]{
  \resizebox*{0.4\textwidth}{0.2\textheight}{\includegraphics{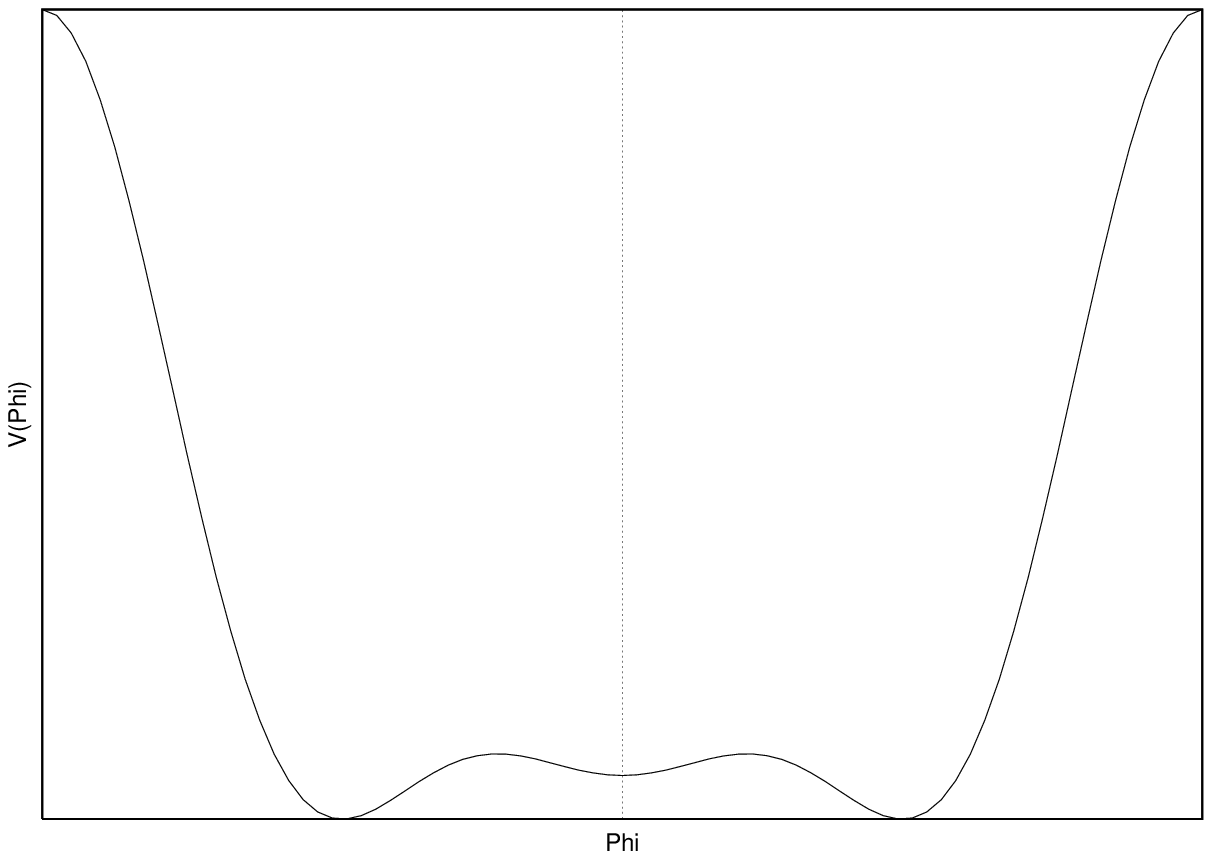}}
 }\\
 \subfigure[$\lambda=\frac{13}{8}\mu$]{
  \resizebox*{0.4\textwidth}{0.2\textheight}{\includegraphics{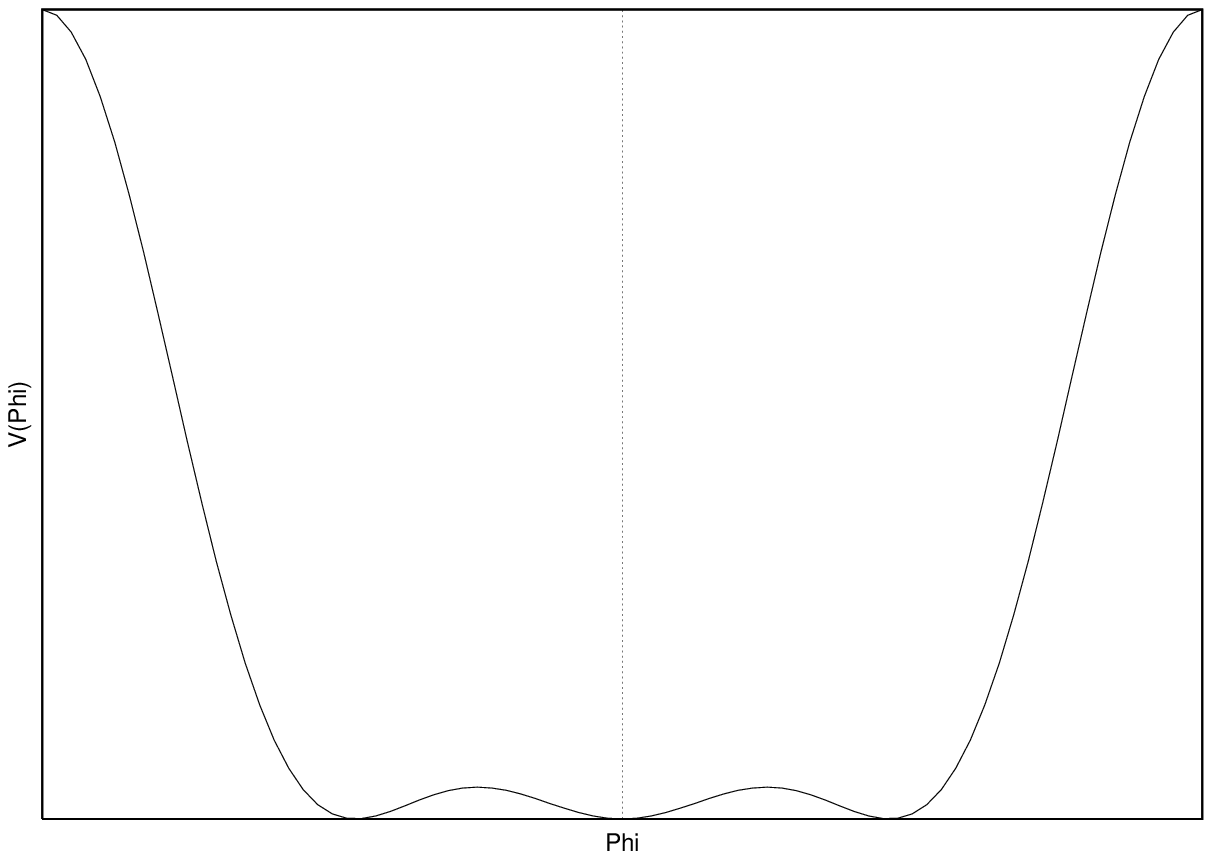}}
 }\\
 \subfigure[$\frac{13}{8}\mu<\lambda<\frac{61}{30}\mu$]{
  \resizebox*{0.4\textwidth}{0.2\textheight}{\includegraphics{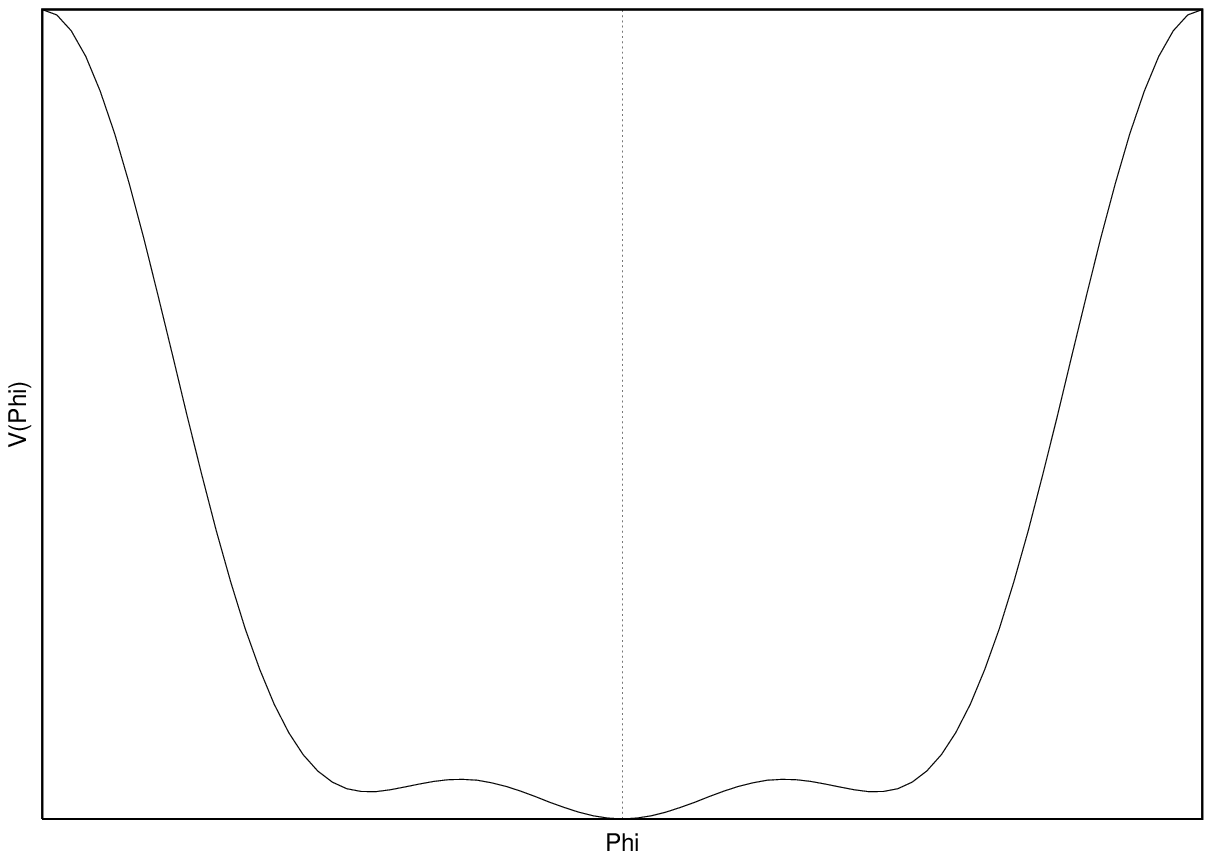}}
 }
 \subfigure[$\frac{61}{30}\mu<\lambda$]{
  \resizebox*{0.4\textwidth}{0.2\textheight}{\includegraphics{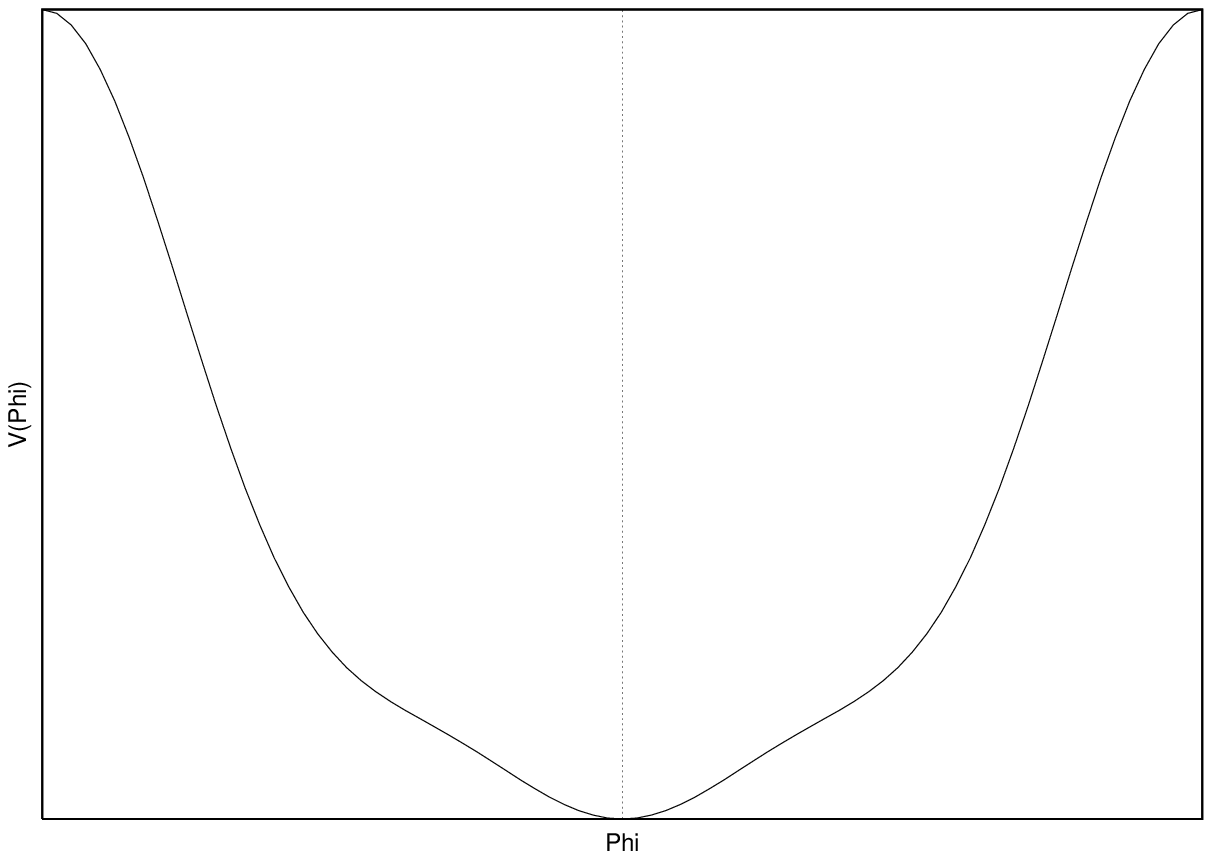}}
 }
\end{center}
\caption{Illustrating a first order transition with $\kappa=-2/5\mu$
and $\nu=0$}
\label{fig:firstordpot}
\end{figure}

Finally we present the characteristic picture of a first order
transition (see Figure \ref{fig:firstordpot}). Since we disagree with
the claim made by Fabrizio et al. that $\sin \frac{3\beta}{2}\Phi$
cannot induce a first order transition, we present a case in which it
does. We fix the following values of the parameters
\begin{equation}
\kappa=-\frac{2}{5}\mu\quad ,\quad \nu=0\,,
\end{equation}
and let $\lambda$ vary from $0$ upwards. At $\lambda=0$, we have only
the two vacua in which ${\mathbb Z}_2$ is spontaneously broken. At
$\lambda=\frac{3}{5}\mu$ the symmetry preserving vacuum appears as a metastable
state. For $\lambda=\frac{13}{8}\mu$ it becomes degenerate with the
symmetry breaking vacua: this is the so-called coexistence point where
the three vacuum states exist simultaneously in the theory. Then for
$\lambda>\frac{13}{8}\mu$ the symmetric ground state becomes the true
vacuum and the two other are metastable ones which disappear at
$\lambda=\frac{61}{30}\mu$.

\subsection{Signatures of 1st and 2nd order phase transitions in finite volume}
\label{signatures}
In order to establish the nature of the phase transition we shall
examine the spectrum of the model in a finite volume. To prepare the
ground we now discuss the signatures of both types of phase transitions
in the finite volume spectrum.

\subsubsection{2nd order phase transitions}

For definiteness we specify the symmetry of the transition to be
${\mathbb Z}_2$, although the discussion can be easily generalized. In
the broken phase, there are two degenerate vacua in infinite
volume. For finite $L$ the degeneracy is lifted by tunneling effects
through the barrier separating them and the lowest lying becomes the
ground state, while the other one is the first excited state. The
split between the ground states vanishes exponentially as
$L\,\rightarrow\,\infty$ because the height of the barrier is
proportional to $L$. For $L\,\rightarrow\,0$ the two ground states
tend to some eigenstates of the conformal Hamiltonian and therefore
the split is proportional to $L^{-1}$.

In the unbroken phase there is only one vacuum. The first excited state
is generally some massive one-particle state and has a finite gap $\mathcal{M}$
over the vacuum. Therefore the difference between the ground state and
the first excited state tends to $\mathcal{M}$ as $L\,\rightarrow\,\infty$ in
the same way as indicated in eqn. (\ref{mass_as}). 

Let us call the parameter in which the transition happens $\eta_{\text{crit}}$, in
analogy with the DSG model. Then the characteristic behaviour can be
summarized if we look at the mass gap $\mathcal{M}$ as a function of $\eta$ and
suppose the unbroken phase is $\eta >\eta_{\text{crit}}$ while the broken one is
$\eta <\eta_{\text{crit}}$. We have the following behaviour
\begin{eqnarray}
\mathcal{M}(\eta)&=&0\quad ,\quad \eta <\eta_{\text{crit}} \nonumber \\
\mathcal{M}(\eta)&>&0\quad ,\quad \eta>\eta_{\text{crit}}\,.
\end{eqnarray}
At exactly $\eta=\eta_{\text{crit}}$ by changing $L$ from $0$ to $\infty$ we
interpolate between the UV $c=1$ CFT and an IR fixed point one, which
consists of an IR CFT and a massive one. Therefore  
  certain states in the spectrum have the
asymptotic (large $L$) behaviour
\begin{equation}\label{e2ndpht}
E_{\Psi}(L)-E_{0}(L)=\frac{2\pi\left(\Delta^+_{\text{IR}}+\Delta^-_{\text{IR}}\right)}{L}
+O\left(L^{-1-\epsilon}\right)\quad
,\quad \epsilon>0\,,
\end{equation}
where the subscript $0$ denotes the ground state (the lowest lying
state in finite volume). These states fall into conformal families of
the infrared fixed point characterizing the transition with conformal
dimensions $\Delta^{\pm}_{\text{IR}}$. As $L\,\rightarrow\,\infty$
they decouple from the rest of the spectrum which remains massive. The
identification of such states allows the extraction of the
characteristics of the infrared fixed point CFT as we shall see
shortly.

\subsubsection{1st order phase transition}
\label{signatures_firstorder}

A first order phase transition is always characterized by the
existence of metastable states. As we have seen in Subsection
\ref{subsec:DSG2}, the existence of such states in finite volume
implies that the space of states separates into excitations over the
true ground state and those over the metastable ones which we called
``runaway'' states. The two sets of states have in general different
bulk terms which coincide only when one tunes the coupling $\eta$ to
the coexistence point $\eta_{\text{coex}}$. At exactly this point the
three vacua have identical energy densities and passing through this
value of $\eta$ the symmetry breaking pair of vacua interchanges with
the symmetric one. This is a very characteristic behaviour which one
would think is very easy to distinguish in the finite volume spectrum.

However, first order phase transitions have a characteristic strength
and as is well-known, weak first order phase transitions are hard to
distinguish from second order ones. To see this more concretely, let
us denote the three vacua by $\Phi_0$ for the symmetric one and
$\Phi_\pm$ for the symmetry breaking ones, respectively. There are two
other interesting points in $\eta$: the lower critical value
$\eta_{\text{lcrit}}$ where $\Phi_0$ appears, and the higher critical
value $\eta_{\text{hcrit}}$ where $\Phi_\pm$ disappear as we increase
$\eta$. In the case of a second order phase transition
\begin{equation}
\eta_{\text{lcrit}}=\eta_{\text{hcrit}}\,.
\end{equation}
If the difference $\eta_{\text{hcrit}}-\eta_{\text{lcrit}}$ is small,
then when tuning $\eta$ numerically (or in an experiment) one may miss
the existence of the metastable vacua. Similarly, one can look at the
difference between the vacuum energy densities
$V\left(\Phi_0\right)-V\left(\Phi_\pm\right)$ at the lower and upper
critical points. If it is small, then to detect the fact that the
space of states separated into two sets of states with different
linear terms in their energies one needs to go to very large values of
the volume where truncation errors in TCSA become large. Therefore it
is clear that one can never eventually distinguish a sufficiently weak
first order transition from a second order one, neither numerically
nor experimentally. An exact solution of the model would of course
disentangle the problem, but given that the theory is not integrable
this is too much to hope for.

\section{The phase diagram in the case $ \frac{\alpha }{\beta }=\frac{1}{2}$,
$ \delta =\frac{\pi }{2}$}
\label{phasediagram}

In this section we use TCSA to verify the classical predictions for
the phase transition, and to clarify the order and the universality
class of the transition. In order to check for the transition we need
TCSA data over a wide range of the volume parameter $l$. Therefore, in
order for TCSA to be useful, we have to restrict our investigations to
the domain $\beta^2<4\pi$ where the method does not have UV
divergences (see the discussion at the end of Subsection
\ref{subsec:tcsa_ham}). It turned out that even in this range we cannot
attain the necessary precision everywhere and to get a convergent
enough numerical method we had to restrict $\beta^2<8\pi/3$. For the
$\eta=0$ sine-Gordon theory, this corresponds to a regime with two or
more breathers in the spectrum.

\begin{figure}
\begin{center}
 \subfigure[$\eta=0.6$]{
  \resizebox*{0.4\textwidth}{0.2\textheight}{\includegraphics{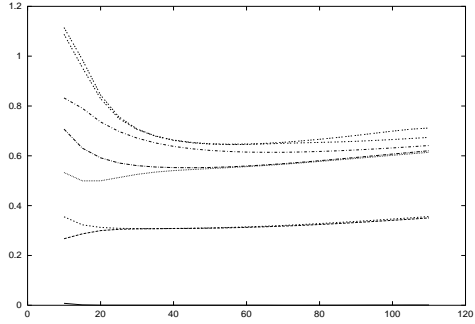}}
 }
 \subfigure[$\eta=0.7$]{
  \resizebox*{0.4\textwidth}{0.2\textheight}{\includegraphics{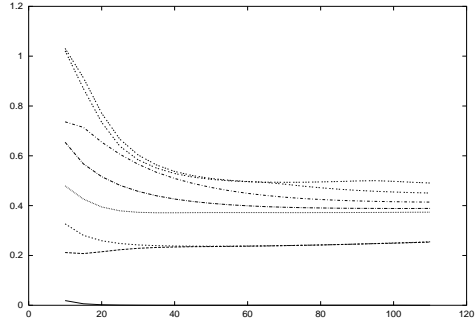}}
 }\\
 \subfigure[$\eta=0.8$]{
  \resizebox*{0.4\textwidth}{0.2\textheight}{\includegraphics{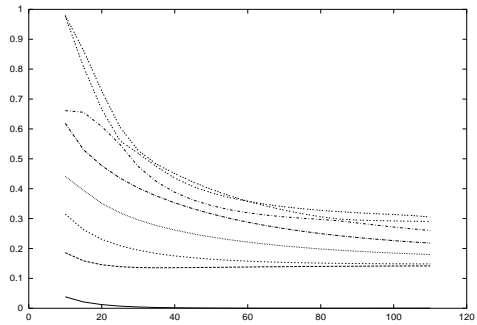}}
 }\\
\subfigure[$\eta=0.84$]{
  \resizebox*{0.4\textwidth}{0.2\textheight}{\includegraphics{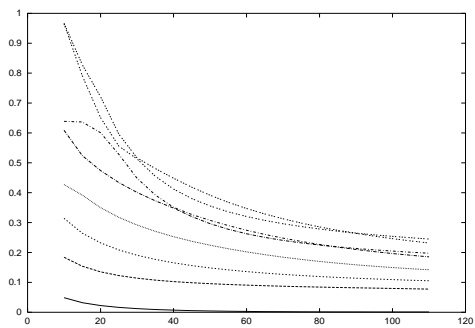}}
 }
 \subfigure[$\eta=0.9$]{
  \resizebox*{0.4\textwidth}{0.2\textheight}{\includegraphics{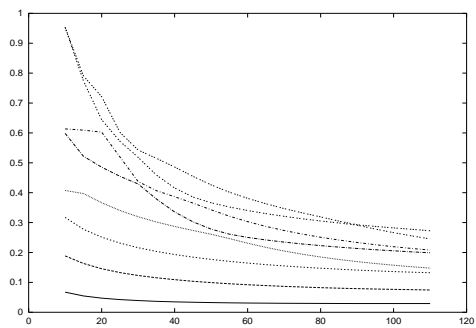}}
 }
\end{center}
\caption{Change of the spectrum as $\eta$ varies from $0$ to $1$ in
$\text{DSG}_2^{\eta}$ for $\beta=8\sqrt{\pi}/7$ and $\delta=\pi/2$. The
energies are normalized by subtracting the ground state
contribution. The plots show the first $8$ states.}
\label{fig:mvve}
\end{figure}

To examine the nature of the phase transition we ran the $\eta$ \lq
movie' with $\delta=\pi/2$ at various values of $\beta$. The results
show no trace of a first order phase transition - a case is depicted
on Figure \ref{fig:mvve}.  Indeed, for the case shown
($\beta=8\sqrt{\pi}/7$) the spectra can be classified into three
classes: (1) for $\eta <0.8$ one can clearly see the doubly degenerate
ground states and the also degenerate first massive states above them;
(2) for $0.9<\eta <1$ the spectrum is massive, but there are no
degeneracies among the lowest lying states and (3) in the transition
domain, for $\eta\sim 0.84$, the structure of the spectrum changes,
there are no lines tending to a constant, instead they decrease in the
whole volume range and the degeneracies of the ground states and the
first excited states are removed. Since at no place could we see the
linearly rising set of runaway states that would correspond to the
presence of metastable vacua with a different bulk energy, we analyze
the data by looking for a 2nd order phase transition at some
$\eta=\eta_{\text{crit}}$. (See however the remarks at the end of
Subsection \ref{signatures_firstorder}).

\subsection{How do we find $ \eta =\eta _{\text{crit}}$?}

We determined the critical value of $\eta$ in the following way: 
we tuned $\eta$ in the transition
region by looking for whether the $\epsilon_2(l)-\epsilon_{\rm
vac}(l)$ difference of the second excited state and the ground state
continues to decrease along the (in between enlarged) complete $l$
range. The meaning of this criterion is clear: this difference
describes both at $\eta=0$ and at $\eta=1$ the mass of some low-lying
breather of the limiting SG theories; thus its continuous decrease at
a specific $\eta_{\text{crit}}$ implies that this breather became massless.
Since this happens only at this particular value of $\eta$, 
we think that this criterion is better than the one based on the appearance
of a gap between the two ground states (as the gap is there for all
$\eta>\eta_{\text{crit}}$). Nevertheless, for consistency, at $\eta_{\text{crit}}$ chosen, 
$\epsilon_1(l)-\epsilon_{\rm
vac}(l)$ should decay slower than exponentially with $l$; in fact, recalling
eqn. (\ref{e2ndpht}), it should be $O(1/l)$ for large $l$.   

\begin{figure}
{\par\centering \includegraphics{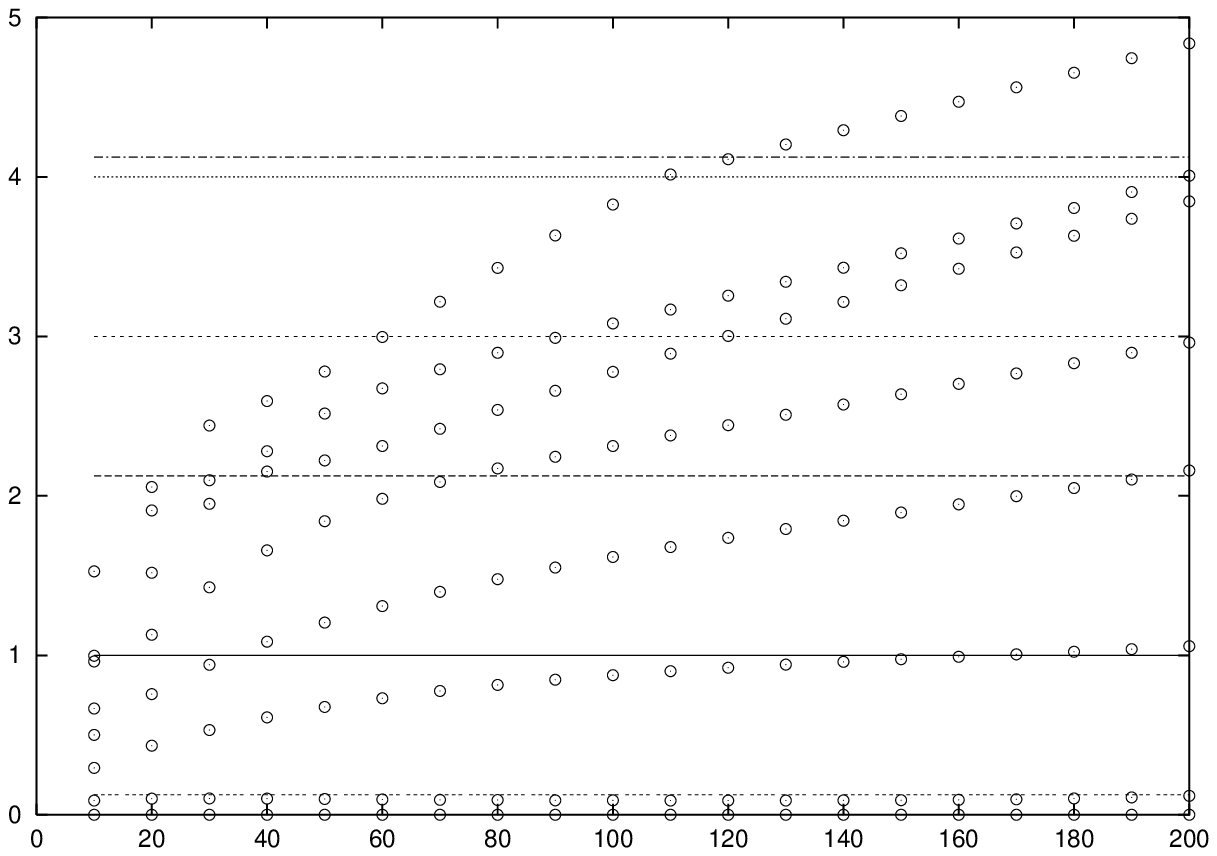} \par}

\caption{$[\epsilon_i(l)-\epsilon_0(l)]\, l/(2\pi )$ at
$\beta=8\sqrt{\pi}/7$ and $\eta=0.866$}
\label{fig:ispl1}
\end{figure}

\begin{figure}
{\par\centering \includegraphics{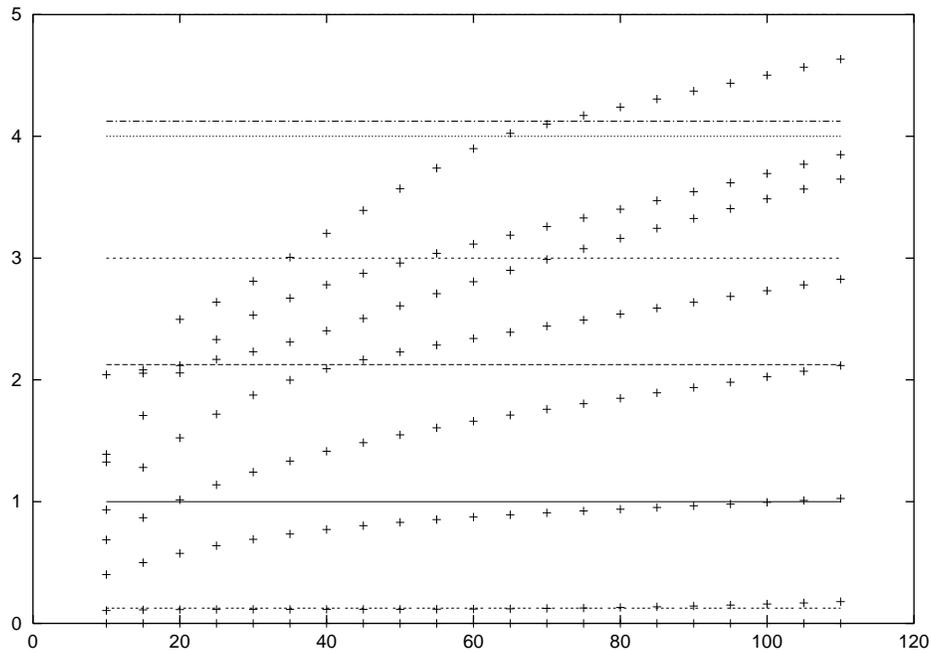} \par}

\caption{$[\epsilon_i(l)-\epsilon_0(l)]\, l/(2\pi )$ at
$\beta=4\sqrt{\pi}/3$ and $\eta=0.850$}
\label{fig:ispl2}
\end{figure}
\begin{figure}
{\par\centering \includegraphics{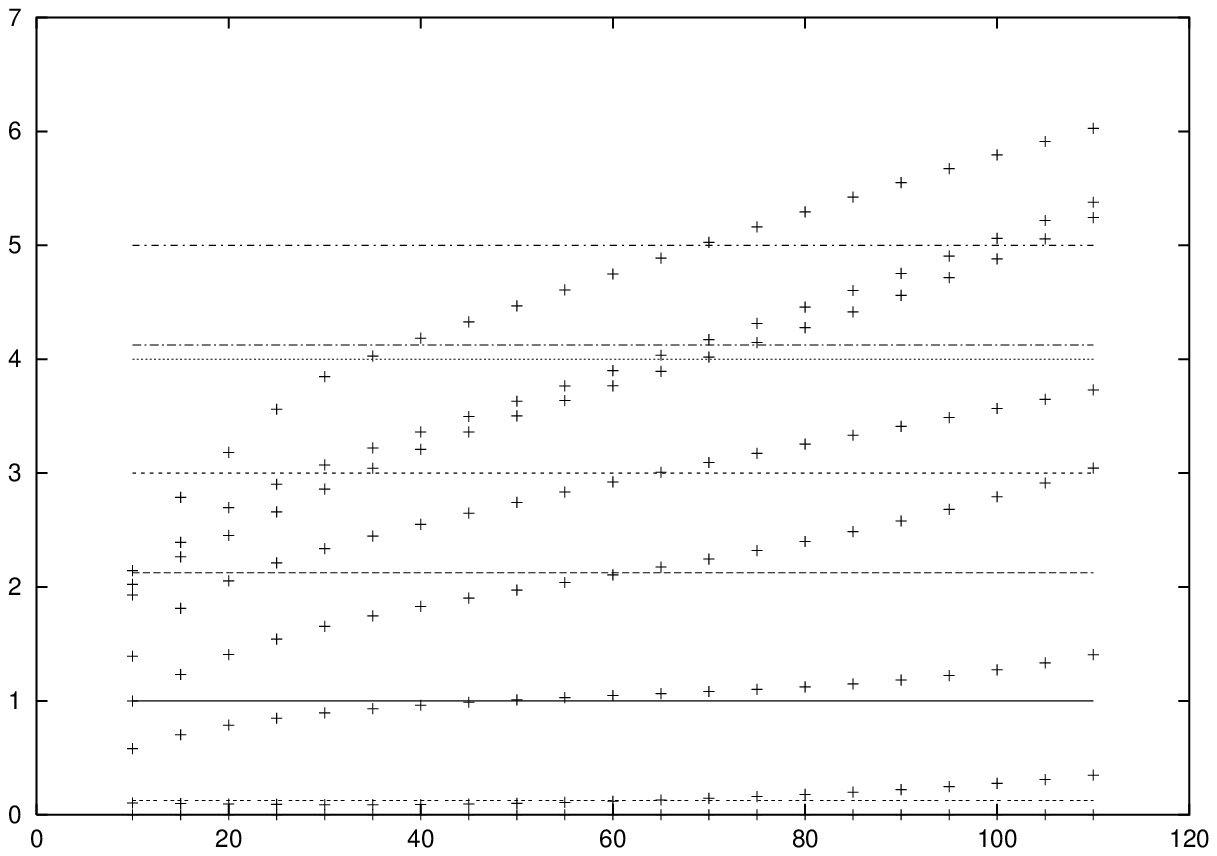} \par}

\caption{$[\epsilon_i(l)-\epsilon_0(l)]\, l/(2\pi )$ at
$\beta=8\sqrt{\pi}/5$ and $\eta=0.838$}
\label{fig:ispl3}
\end{figure}
The result of this search is shown on
Figures \ref{fig:ispl1}-\ref{fig:ispl3} for three different values of
$\beta$. All three plots were taken at the \lq best' $\eta$ values
according to our criterion. To separate more clearly the various  
$\epsilon_i(l)-\epsilon_0(l)$ as functions of $l$ on all three plots
we show these differences multiplied by $l/(2\pi)$. (This also makes
them more suitable for the subsequent analysis). 
Here $\epsilon_0(l)$
denotes the lowest eigenvalue (vacuum energy) and $i$ runs from $0$ to
$7$. The horizontal lines on the plots correspond to the possible
values of $\Delta^+_{\rm IR}+\Delta^-_{\rm IR}$ if we assume that the
IR CFT is the $c=1/2$ Ising model (see eqn. (\ref{e2ndpht})), when the
spectrum of $\Delta^+_{\rm IR}+\Delta^-_{\rm IR}$ is determined by the
Ising primary fields and their descendants.

In all three cases we see that the first two lines - that correspond
to the first two excited states above the ground state - qualitatively fit rather
well to the Ising model's prediction. For the higher excited states
the agreement is not so spectacular but even for them the sequence of lines
and the absence of degeneracies (which on the Ising side follow from
the null vector pattern) match that of the Ising prediction. For the
two smaller $\beta$ values it is clear that also some of the higher
excited states would reach their corresponding plateaus  
by going to even higher volumes\footnote{These plots show that the
higher the value of $\beta$ the larger are the TCSA errors. The
spectra were computed by using a conformal energy cut $E_{cut}\sim 15-
16$ which resulted in $4.5-5.5\times 10^3$ states in the UV conform
Hilbert space.}.  

To make this agreement more
quantitative we have to describe two types of effects. On the one hand
we need the spectrum of states on the
critical trajectory (i.e. 
at $\eta=\eta_{\text{crit}}$) in a finite volume more precisely than in
eqn. (\ref{e2ndpht}),  
and on the other
we need a clear, quantitative picture of the deviations from
criticality since our best $\eta$ may not equal $\eta_{\text{crit}}$ (and 
TCSA errors may also be involved). 

\subsection{Theoretical predictions for
the form of the spectra}

 We describe these effects
by assuming that the IR CFT is the critical Ising model, i.e. 
 by using the Hilbert space and operator content of
the Ising model. 

Let us concentrate first on the finite volume corrections at
$\eta=\eta_{\text{crit}}$. In this case when changing $l$ from zero to
infinity we interpolate from the UV conformal fixed point to an IR one,
at least if we neglect the coupling between the massive and massless
modes in the IR. The massive and massless modes decouple exactly at
$l=\infty$ and we shall call the subspace of states asymptotically
corresponding to the massless ones the \textit{scaling sector}. Since
one can flow into a fixed point only from an irrelevant direction, in
this approximation the interpolating theory may be considered as an
irrelevant perturbation of the conformal Ising model with (the non
renormalizable) action
\begin{equation}
{\cal A}={\cal A}_{\rm Ising}+g\int d^2z(\psi (z,\bar{z})+{\rm higher\
 dimensional\ fields})\,.
\end{equation}
The various directions correspond to various choices of $\psi
(z,\bar{z})$, which in turn are expressed in terms of the irrelevant
(spinless and non vanishing) descendants of the primary fields. Since
the DSG$^\eta_2(\beta ,\pi /2)$ model exhibits the ${\mathbb Z}_2$ 
symmetry for all values of $\eta$, $\psi (z,\bar{z})$ may contain only
the descendants of the energy operator $\partial
\bar{\partial}\epsilon (z, \bar{z})+$higher dimensional fields and of
the identity operator $T\bar{T}+$higher dimensional fields (where
$\bar{T}(T)$ is the (anti)holomorphic component of the stress energy
tensor of the Ising model). Recalling the generic form of the TCSA
Hamiltonian,   
\begin{equation}
H=\frac{2\pi }{L}\left(
L_{0}+\bar{L}_{0}-\frac{c}{12}+\frac{g L^{2-2\Delta _{\psi
}}}{(2\pi )^{1-2\Delta _{\psi }}}\hat{\psi} (1,1)\right)\,,
\end{equation}
we see that for large but finite $l$ the leading corrections to
$E_\Psi (L)-E_0(L)$ come from the least irrelevant operator:
\begin{equation}
\epsilon_\Psi(l)-\epsilon_0(l)=\frac{2\pi}{l}(\Delta^+_{\rm
IR}+\Delta^-_{\rm IR})+A_\Psi l^{1-2\Delta_\psi}+\dots 
\end{equation}
(Here the constant $A$ may depend on the state in question). Thus, in
our case, the leading correction takes the form 
\begin{equation}
\epsilon_\Psi(l)-\epsilon_0(l)=\frac{2\pi}{l}(\Delta^+_{\rm
IR}+\Delta^-_{\rm IR})+A_\Psi l^{-2}+\dots 
\end{equation}
if $\psi$ is given by the derivative of the energy operator. (If this
term were absent then $T\bar{T}$ would give a correction
$\tilde{A}_\Psi l^{-3}$). 

Numerically we can probably never tune $\eta$ exactly to $\eta_{\rm
c}$, and even if we could the TCSA errors -- coming from truncating the
Hilbert space -- would drive away from the critical trajectory
connecting the UV and IR conformal theories. From the side of the
Ising model this means that some relevant perturbation is also
switched on, and we should also take its effects into account when
discussing the large $l$ spectrum. In the Ising model there is only
one relevant perturbation compatible with the  ${\mathbb Z}_2$ symmetry, namely
when the perturbing operator is $\epsilon (z,\bar{z})$. The presence of
a relevant perturbation $\gamma \hat{\epsilon }(1,1)$ in the TCSA
Hamiltonian leads to a correction $B_\Psi$ or $B_\Psi +C_\Psi l$ to 
$\epsilon_\Psi(l)-\epsilon_0(l)$\footnote{The second term seems to be
dominant over the first one, however its coefficient -- which in a pCFT
framework is of second order -- in general is much smaller than the
constant, thus for a large but finite range of $l$-s they may be
equally important.}. Thus, summarizing, we expect, that in a large but
finite volume range the energy spectrum can be described by the
following:
\begin{equation}
\epsilon_i(l)-\epsilon_0(l)=B_i+\frac{2\pi}{l}D_i+A_i l^{-2}+C_il+\dots
\label{function_to_fit}
\end{equation}
In the following we fit the TCSA data using this expression. We keep
all these terms since we are at an intermediate volume range, where
all of them can be equally important. A little bit more abstractly we
can say that on the \lq\lq best $\eta$'' trajectory we come into the
vicinity of the IR fixed point, where the features of this CFT determine
the energy spectrum, but eventually, because of the relevant
perturbation, we are driven away and flow into a massive
theory. Nevertheless, if in the intermediate volume range these
expressions describe the spectra well, -- in particular if the
measured values of $B_i$ (or $B_i$ and $C_i$) are small and that of
$D_i$ are similar to the ones predicted by the Ising model -- then
this confirms the presence of a 2nd order phase transition in the
Ising universality class.  We call this intermediate volume range the
\textit{scaling region} which may (and in fact does, see Figures
\ref{fig:ispl1}, \ref{fig:ispl2} and \ref{fig:ispl3}) change from
state to state.

Let us pause for a short discussion of the plateaus apparent in
Figures \ref{fig:ispl1}, \ref{fig:ispl2} and \ref{fig:ispl3}. Note
that if we had the exact finite volume spectrum (i.e. without
truncation errors) then the scaling functions
\begin{equation}
\frac{l}{2\pi}\left(\epsilon_i(l)-\epsilon_0(l)\right)
\end{equation}
would tend to a constant as $l\,\rightarrow\,\infty$ for
$\eta=\eta_{\text{crit}}$. In this sense the IR fixed point lies
infinitely far away and the above functions approach asymptotic
``plateaus'' at $l=\infty$. 

In the real TCSA situation, the scaling region corresponds to the
plateaus in the numerical scaling functions. Tuning
$\eta$ closer and closer to the critical value these plateaus become
longer and flatter and in addition they move to higher and higher
values of the volume $l$.

Experience with TCSA for perturbed $c=1$ CFTs shows that
truncation errors generally decrease if one decreases $\beta$ (or
equivalently, increases the compactification radius of the conformal boson
field). As a result of the above considerations, we expect that the
plateaus are longer and flatter for smaller $\beta$ and that they
start at higher values of $l$. This is in fact what we observed in
practice (see Figures \ref{fig:ispl1}, \ref{fig:ispl2} and
\ref{fig:ispl3}).

One can also make a theoretical prediction for the form of the energy
$\epsilon_0(l)$ of the ground state as well. Using well-known facts
about the off-critical Ising model \cite{klassen_melzer1} and
arguments similar to the above one expects the following behaviour in
the scaling region
\begin{equation}
\epsilon_0(l)=-\frac{\pi c}{6 l}+a_0 l^{-2}+b_0+c_0l+\tilde{c}_0l\log
l +\dots 
\label{vacuum_to_fit}
\end{equation}
where $c=1/2$ is the central charge of the Ising model and the
logarithmic term appears as a result of a ``resonance'' between the
linear (bulk) vacuum energy contribution and the perturbative
corrections coming from the relevant operator $\epsilon$. The
term $\tilde{c}_0l\log l$ is universal for all the excited states in
the scaling sector and so disappears from energy differences like  
(\ref{function_to_fit}).

\subsection{Numerical results}

\subsubsection{Scaling dimensions and UV-IR operator correspondence}

One can fit the function (\ref{function_to_fit}) to the TCSA data in
the scaling regions plotted in Figures \ref{fig:ispl1},
\ref{fig:ispl2} and \ref{fig:ispl3}. The results of the fits are shown
in table \ref{tab:fit}.

\begin{table}
\begin{center}
\subtable[$\beta=8\sqrt{\pi}/7$, $\eta=0.866$]{
\begin{tabular}{|l|c|c|c|c|}
\hline
State & $D_i$ & $A_i$ & $B_i$ & $C_i$ \\
\hline
\hline
$i=1$ & $0.125 \pm 0.002$ & $-1.2 \pm 0.3$ & $-0.0034\pm 0.0001$ & $1.45\times 10^{-5}\pm 6\times 10^{-7}$ \\
\hline
$i=2$ & $1.04 \pm 0.03$ & $-128 \pm 8$ & $0.0017 \pm 0.0018$ &
$9\times 10^{-6}\pm 5\times 10^{-6}$\\
\hline
\end{tabular}
}
\subtable[$\beta=4\sqrt{\pi}/3$, $\eta=0.850$]{
\begin{tabular}{|l|c|c|c|c|}
\hline
State & $D_i$ & $A_i$ & $B_i$ & $C_i$ \\
\hline
\hline
$i=1$ & $0.1312 \pm 0.0009$ & $-1.35 \pm 0.03$ & $-0.0061\pm 0.0003$ & $2.6\times 10^{-5}\pm 6\times 10^{-6}$ \\
\hline
$i=2$ & $1.03 \pm 0.02$ & $-74 \pm 3$ & $0.005 \pm 0.002$ &
$1.4\times 10^{-5}\pm 9\times 10^{-6}$\\
\hline
\end{tabular}
}
\subtable[$\beta=8\sqrt{\pi}/5$, $\eta=0.838$]{
\begin{tabular}{|l|c|c|c|c|}
\hline
State & $D_i$ & $A_i$ & $B_i$ & $C_i$ \\
\hline
\hline
$i=1$ & $0.1455 \pm 0.0005$ & $-0.89 \pm 0.05$ & $-0.0145\pm 0.0005$ & $2.24\times 10^{-4}\pm 9\times 10^{-6}$ \\
\hline
$i=2$ & $1.09 \pm 0.02$ & $-40 \pm 2$ & $-0.0015 \pm 0.0031$ &
$1.1\times 10^{-4}\pm 2\times 10^{-5}$\\
\hline
\end{tabular}
}
\caption{The results of fitting (\ref{function_to_fit}) to the first
two excited states for various values of $\beta$ at the estimated
critical value of $\eta$}
\label{tab:fit}
\end{center}
\end{table}

From the conformal data of the Ising fixed point we know that the
first excited state should correspond to the spin operator $\sigma$
with conformal dimensions $\Delta^\pm=1/16$, while the second to the
energy operator $\epsilon$ with conformal dimensions $\Delta^\pm=1/2$.
Therefore we expect 
\begin{equation}
D_1=\frac{1}{8}=0.125\quad ,\quad D_2=1\,.
\end{equation}
The data presented fit quite well with these predictions. Note that
the agreement becomes better for smaller $\beta$ where the TCSA errors
are smaller. We mention that the errors presented in the tables come
from the fit procedure and do not contain the truncation errors which
are generally much larger.

Another piece of information is the UV-IR operator correspondence. In
TCSA we know the operator in the UV theory which creates the
excited states. By identifying these states in the language of the IR
theory, one can establish a correspondence between the UV and the IR
operator content. In particular, the first and the second excited
states are created by the operators
\begin{equation}
\cos\frac{\beta}{2}\Phi\quad\text{and}\quad\sin\frac{\beta}{2}\Phi
\end{equation}
in the UV, respectively. In the IR these states flow to ones
corresponding to the Ising operators $\sigma$ and $\epsilon$. This is
in full accordance with their parity under the ${\mathbb Z}_2$
symmetry (\ref{Z2}) and the fact that $\sigma$ is an odd, while
$\epsilon$ is an even operator in the Ising model. In addition, both
the operator $\sin\frac{\beta}{2}\Phi$ in the $\text{DSG}_2^\eta$
model and the operator $\epsilon$ in the Ising case correspond to the
``temperature'' parameter deforming the model away from criticality.
Furthermore, this conclusion also agrees with the conjecture made by Fabrizio et
al. \cite{nersesyan}.

One could also try to extract the central charge $c$ using the
predicted form of the vacuum energy function (\ref{vacuum_to_fit}). It
turns out, however, that the fit is not sensitive enough to the value
of $c$ --- in a least-squares algorithm the main contribution to
$\chi^2$ comes from the linear and $l\log l$ terms and $c$ can
only be determined with $30-40 \%$ accuracy, within which we find
agreement with the Ising value $c=1/2$.

\subsubsection{Analysis of the phase diagram}

Concerning the phase diagram, our first observation is that all
evidence points to a second-order phase transition for all values of
$\beta$, which happens at some critical value
$\eta_{\text{crit}}(\beta)$ dependent on $\beta$. As we already
discussed in Subsection \ref{signatures}, there is always a
possibility of a sufficiently weak first order transition. The only
statement one can make is that the phase transition is second order up
to the precision within which the coefficients $B_i$ and $C_i$ in
($\ref{function_to_fit}$) vanish at the (approximate value of the)
critical coupling. Given the data in table \ref{tab:fit} and the fact
that the normal precision in TCSA is of the order $10^{-2}-10^{-3}$
for the values of the volumes used (this can be estimated
e.g. calculating an integrable case when the exact values of the
excited state energies are known) we can state that \emph{the phase
transition is second order up to the precision of the TCSA method}.

As we have already seen the second order nature of the phase
transition is controlled by inequalities like
(\ref{2ndordercond1}). Since a very weakly first order
transition requires a very weak breaking of these inequalities this
in turn strongly constrains the parameters in the effective potential.
However, the coefficients in the effective potential are
uniquely determined in terms of $\lambda$ and $\mu$ and therefore such a fine
tuning is highly unlikely to happen. 

It is expected that for $32\pi/9<\beta^2<8\pi$ the transition is always
second order since then all the possible corrections
(\ref{higher_ops}) to the effective potential correspond to irrelevant
operators. Therefore it is only necessary to consider the case  
$0<\beta^2<32\pi/9$. As TCSA is convergent for $\beta^2<4\pi$ it can be
used to examine this range of couplings and all the values of $\beta$
were from this range. Unfortunately, TCSA does not converge very well
for $\beta^2>8\pi/3$, so there is a gap in the values of $\beta^2$ up to
$32\pi/9$ which is not accessible.

One also expects a second order transition around $\beta=0$ since in
this case the theory is semiclassical and the contributions to the
effective potential are expected to be very small. We also know the
exact value (\ref{etac_semicl})

\begin{equation}
\eta_{\text{crit}}(\beta =0)=\frac{16}{17}=0.941\dots\ .
\end{equation}

Therefore we conjecture that \emph{the
phase transition in $\text{DSG}_2^\eta$ is second-order for all values of
$\beta$ in the exact theory too}.

\begin{figure}[tbp]
\begin{center}
\input{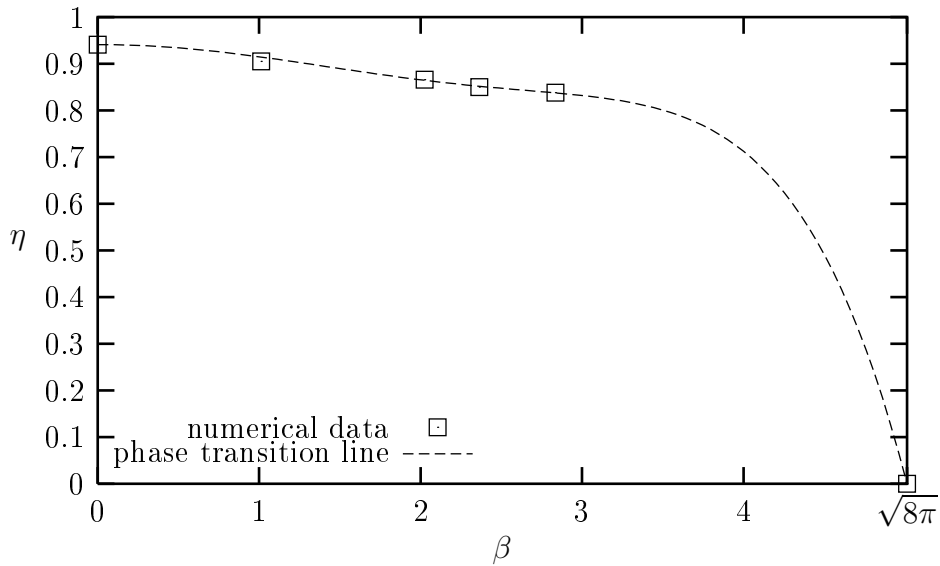}
\end{center}
\caption{The phase diagram}
\label{fig:etac}
\end{figure}

One can prepare a phase diagram in the $(\eta,\beta)$ plane. For
$\eta=0$ the theory is in the broken symmetry phase, while for
$\eta=1$ the ${\mathbb Z}_2$ symmetry is restored. The phase
transition line is described by a function
$\eta=\eta_{\text{crit}}(\beta)$. One can also calculate the value of
this function at the point $\beta=\sqrt{8\pi}$. For
$\beta>\sqrt{8\pi}$ the term $\cos\beta\Phi$ becomes irrelevant and
the theory is always in a symmetric phase. Therefore we expect that 
\begin{equation}
\eta_{\text{crit}}\left(\beta =\sqrt{8\pi}\right)=0\ .
\end{equation}
Since the transformation $\beta\,\rightarrow\,-\beta$ can be
compensated by a field redefinition $\Phi\,\rightarrow\,-\Phi$, one
also has 
\begin{equation}
\eta_{\text{crit}}\left(-\beta\right)=\eta_{\text{crit}}\left(\beta\right)\ .
\end{equation}
Figure \ref{fig:etac} shows the phase diagram of the theory. The
numerical data are the TCSA results for $\eta_{\text{crit}}$ at
particular values of $\beta$ (in addition to the ones in Table \ref{tab:fit},
there is also a point with $\beta=4\sqrt{\pi}/7$ where
the measured critical coupling is $\eta_{\text{crit}}\approx 0.916\pm 0.002$). The
continuous line is an (even) polynomial interpolating the numerical
data together with the two known values at $\beta=0$ and
$\sqrt{8\pi}$, which is shown in order to get an idea of the phase
transition line. Below the transition line the theory is in the broken
phase, while above the line the ${\mathbb Z}_2$ symmetry is restored.

\section{Conclusions}
\label{sec:conclusions}

In the present paper we examined the two-frequency sine-Gordon model, using a
combination of various (analytic and numerical) approaches. Using form
factor perturbation theory (FFPT) on the one hand and truncated
conformal space approach (TCSA) on the other, we first verified the
consistency of the two by comparing the results in the two
perturbative regimes. Based on the agreement between the two methods,
we continued to use TCSA in the nonperturbative regime.

Then we set up a framework for examining phase transition using the
finite volume spectrum extracted from TCSA. We discussed how to
distinguish between first and second order transitions, and in
particular how to extract the characteristic quantities of
an eventual IR fixed point. To our knowledge this is the first case
when TCSA was used for locating and characterizing a nontrivial fixed
point. We would like to mention that TCSA was used before in the
similar context of integrable massless flows in order to check
results from Thermodynamics Bethe Ansatz (TBA) close to the UV CFT
(small values of the volume $l$) \cite{klassen_melzer2}. Here, on the contrary, the TCSA
method was used for finding the critical point at which the flow
becomes massless and then analyzing the intermediate volume range in
which the theory is governed by its IR fixed point. In addition, in
our case there are no alternative nonperturbative methods since the
theory is nonintegrable.

Using this framework, we examined the prediction by Delfino and
Mussardo \cite{delfino_mussardo} of a second order phase transition,
in a particular case when the ratio of the two frequencies was
$1/2$. Contrary to a possibility raised by Fabrizio et
al. \cite{nersesyan} we argued that this phase transition is second
order for \emph{all} possible values of the sine-Gordon-like parameter
$\beta$, at least within the precision attainable by TCSA. We examined
the spectrum in detail and showed that the lowest lying states match
the pattern of an Ising type fixed point, both qualitatively and
quantitatively. We also examined the UV-IR operator correspondence and
found an agreement with the conjectures made by Fabrizio et al.

In addition, by a simple mean field analysis we have shown that
contrary to the claims made by Fabrizio et al., \emph{all} lowest order
correction terms to the effective potential are able to induce a first
order transition, although the numerical studies suggest their
coefficients are such that they do not affect the second-order nature
of the transition in this particular case.

In this paper we restricted ourselves to the investigation of the
vacuum sector. It could be interesting to see the effect of the phase
transition in the topologically nontrivial sectors as well. Another
open question is to examine other possible frequency ratios and to
extract the phase diagram for the general case. Yet another direction
for further study is the case of irrational frequency ratios: as we
already discussed in the paper, the perturbative approach (FFPT) is
still applicable, but there are no nonperturbative alternatives at
present. Finally, as the model has potentially interesting condensed
matter applications, it is also a challenge to use the above framework
to extract some quantities of interest in that context.

\vspace{1cm}

\begin{center}\textbf{Acknowledgements}\end{center}
G. Tak\'acs would like to thank G. Mussardo for useful
discussions. G. T. is supported by a PPARC (UK) postdoctoral
fellowship, while Z. B. by an OTKA (Hungary) postdoctoral fellowship
D25517. This research was also supported in part by the Hungarian
Ministry of Education under FKFP 0178/1999 and the Hungarian National
Science Fund (OTKA) T029802/99.

\vspace{1cm}

\appendix

\makeatletter
\renewcommand\theequation{\hbox{\normalsize\Alph{section}.\arabic{equation}}}
\makeatother

\section{Vacuum expectation values and breather form factors in
(folded) sine-Gordon theory \label{ffs_and_vevs}}

Here we list the vacuum expectation values and form factors necessary for performing
the form factor perturbation theory calculations in the main text, using the
results in \cite{exact_vevs, formfactors}. We define our conventions for sine-Gordon
theory with the action
\begin{equation}
\mathcal{A}_{SG}=\int \ud t\int \ud x\left( \frac{1}{2}\partial _{\mu }\Phi \partial ^{\mu }\Phi +\mu \, :\, \cos \beta \Phi \, :\right)\,. \end{equation}
 The vacuum expectation value of the exponential operator
\begin{eqnarray}
V_{a}(x) & = & :\, \exp \left( ia\Phi (x)\right) \, :\\
\mathcal{G}_{a}(\beta ) & = & \big \langle 0\big |V_{a}(0)\big |0\big \rangle 
\end{eqnarray}
is of the form \cite{exact_vevs}
\begin{eqnarray}
\mathcal{G}_{a}(\beta ) & = & \left[ \frac{M\sqrt{\pi }\Gamma \left( \frac{4\pi }{8\pi -\beta ^{2}}\right) }{2\Gamma \left( \frac{\beta ^{2}/2}{8\pi -\beta ^{2}}\right) }\right] \nonumber \\
 & \times  & \exp \left\{ \int ^{\infty }_{0}\frac{\ud t}{t}\left[ \frac{\sinh ^{2}\left( \frac{a\beta }{4\pi }t\right) }{2\sinh \left( \frac{\beta ^{2}}{8\pi }t\right) \sinh \left( t\right) \cosh \left( \left( 1-\frac{\beta ^{2}}{8\pi }\right) t\right) }-\frac{a^{2}}{4\pi }e^{-t}\right] \right\}\,, \label{lz_vev} 
\end{eqnarray}
provided we normalize our operators in the following way:
\begin{equation}
\big \langle 0\big |V_{a}(x)V_{-a}(x')\big |0\big \rangle =\frac{1}{\left| x-x'\right| ^{\frac{a^{2}}{4\pi }}}\quad \text{as}\quad \left| x-x'\right| \, \rightarrow \, 0\: .\end{equation}
Here $ M $ denotes the soliton mass and $ \left| 0\right\rangle  $ is the vacuum
in which the sine-Gordon field $ \Phi  $ has vanishing vacuum expectation
value.

Let us use the following notations for the breather form factors
\begin{equation}
F^{(a)}_{k_{1}\dots k_{n}}\left( \vartheta _{1},\dots ,\vartheta _{n}\right) =\big \langle 0\big |V_{a}(0)\big |B_{k_{1}}\left( \vartheta _{1}\right) \dots B_{k_{n}}\left( \vartheta _{n}\right) \big \rangle ^{in}_{0}\,,\end{equation}
where $ B_{k} $ denotes the $ k $th breather with mass
\begin{equation}
M_{k}=2M\sin \left( \frac{\pi k}{2}p\right) \quad ,\quad p=\frac{\beta ^{2}}{8\pi -\beta ^{2}}\: .\end{equation}
Then we have from \cite{formfactors}
\begin{equation}
\label{ff_11}
F^{(a)}_{11}\left( \vartheta _{1},\vartheta _{2}\right) =-\mathcal{G}_{a}(\beta )\bar{\lambda }^{2}[a]^{2}R\left( \vartheta _{1}-\vartheta _{2}\right)\,, 
\end{equation}
and\\
\begin{eqnarray}
F^{(a)}_{1111}\left( \vartheta _{1},\vartheta _{2},\vartheta _{3},\vartheta _{4}\right)  & = & \mathcal{G}_{a}(\beta )\bar{\lambda }^{4}[a]^{2}\prod _{1\leq k<j\leq 4}R\left( \vartheta _{k}-\vartheta _{j}\right) \times \nonumber \\
 &  & \left\{ [a]^{2}+\left( \sigma _{1}^{2}\sigma _{4}+\sigma _{3}^{2}\right) \prod _{1\leq i<j\leq 4}\left( x_{i}+x_{j}\right) ^{-1}\right\} \label{ff_1111}\,, 
\end{eqnarray}
where we used the following notations
\begin{eqnarray}
 &  & x_{k}=\exp \left( \vartheta _{k}\right) \quad ,\quad [a]=\frac{\sin \left( p\pi \frac{a}{\beta }\right) }{\sin p\pi }\,,\\
 &  & \bar{\lambda }=2\cos \frac{p\pi }{2}\sqrt{\sin \frac{p\pi }{2}}\exp \left( -\int ^{p\pi }_{0}\frac{\ud t}{2\pi }\frac{t}{\sin t}\right)\,, \\
 &  & \sigma _{1}=\sum ^{4}_{i=1}x_{i}\quad ,\quad \sigma _{4}=\prod ^{4}_{i=1}x_{i}\quad ,\quad \sigma _{3}=\sigma _{4}\sum ^{4}_{i=1}x^{-1}_{i}\,,
\end{eqnarray}
and in the strip $ -2\pi +p\pi \leq \Im m\, \vartheta \leq -p\pi  $ (note that
in the attractive regime $ p<1 $) the function $ R(\vartheta ) $ is defined as 
\begin{equation}
\label{minimal_formfactor}
R(\vartheta )=\mathcal{N}\exp \left\{ 8\int \frac{dt}{t}\frac{\sinh (t)\sinh (pt)\sinh ((1+p)t)}{\sinh ^{2}2t}\sinh ^{2}\left[ \left( 1-\frac{i\vartheta }{\pi }\right) t\right] \right\}\,, 
\end{equation}
with
\begin{equation}
\label{Nfactor}
\mathcal{N}=\exp \left\{ 4\int \frac{dt}{t}\frac{\sinh (t)\sinh
(pt)\sinh ((1+p)t)}{\sinh ^{2}2t}\right\}\: . 
\end{equation}
$ R(\vartheta )$ satisfies the useful identity
\begin{equation}
\label{R_funcrel}
R\left( \vartheta \right) R\left( \vartheta \pm i\pi \right) =\frac{\sinh \vartheta }{\sinh \vartheta \mp \sin p\pi }
\end{equation}
which can be used to continue it analytically out of the validity
range of its integral representation.

For the mass correction of the second breather we need the explicit
form of the form factor $ F^{(a)}_{22} $. This is not available in the
literature, but can be computed from $F^{(a)}_{1111}$ using
the bootstrap equation
\begin{eqnarray}
i\, \mathop {\text{Res}}_{\epsilon \, \rightarrow \, 0}\,  & F_{a_{1}\dots a_{n}11}\left( \vartheta _{1},\dots ,\vartheta _{n},\vartheta _{n+1}+i\bar{U}^{1}_{12}-\frac{\epsilon }{2},\vartheta _{n+1}-i\bar{U}^{1}_{12}+\frac{\epsilon }{2}\right) = & \\
 & \Gamma ^{2}_{11}F_{a_{1}\dots a_{n}2}\left( \vartheta _{1},\dots
 ,\vartheta _{n},\vartheta _{n+1}\right)\: .  & 
\end{eqnarray}
The fusion angle $ \bar{U}^{1}_{12} $ and the three-point coupling
$ \Gamma ^{2}_{11} $ are defined from the $ B_{2} $ pole $ U^{2}_{11} $
in the $ B_{1}-B_{1} $ $ S $-matrix:
\begin{equation}
S_{11}(\vartheta )\sim \frac{i\left( \Gamma ^{2}_{11}\right) ^{2}}{\vartheta -iU^{2}_{11}}\quad ,\quad \vartheta \sim iU_{11}^{2}\quad ,\: \text{where}\quad S_{11}(\vartheta )=\frac{\sinh \vartheta +i\sin p\pi }{\sinh \vartheta -i\sin p\pi }\end{equation}
and from the relations
\begin{equation}
\bar{U}^{1}_{12}=\pi -U^{1}_{12}\quad ,\quad U^{2}_{11}+2U^{1}_{12}=2\pi \: .\end{equation}
This way we find
\begin{equation}
U^{2}_{11}=2\bar{U}^{1}_{12}=p\pi \quad ,\quad \left( \Gamma ^{2}_{11}\right) ^{2}=2\tan p\pi \: .\end{equation}
The residue of $ R $ can be calculated from (\ref{R_funcrel})
\begin{equation}
R\left( -ip\pi +\epsilon \right) =-\frac{i}{\epsilon }\tan p\pi \frac{1}{R\left( -i\pi (1+p)\right) }\end{equation}
and finally the $ B_{2}-B_{2} $ form factor is
\begin{eqnarray}
F^{(a)}_{22}(\vartheta _{1},\vartheta _{2}) & = & \left( \Gamma ^{2}_{11}\right) ^{-2}K_{22}\left( \vartheta \right) R\left( \vartheta \right) ^{2}R\left( -\vartheta -ip\pi \right) R\left( -\vartheta +ip\pi \right) \frac{\tan ^{2}\left( p\pi \right) }{R\left( -i\pi (1+p)\right) ^{2}}\: ,\nonumber \\
K_{22}\left( \vartheta \right)  & = & \mathcal{G}_{a}(\beta )\bar{\lambda }^{4}[a]^{2}\left\{ [a]^{2}+\frac{1}{\cosh \left( \vartheta \right) +\cos \left( p\pi \right) }\right\} \quad ,\quad \vartheta =\vartheta _{1}-\vartheta _{2}\: .\label{ff_22} 
\end{eqnarray}
Finally we recall that in the perturbative investigation of the DSG
model we need the vacuum expectation values
and form factors in th sectors over the vacua 
$ \left| k\right\rangle$ characterized by the property
\begin{equation}
\left\langle k\right| \Phi (x)\left| k\right\rangle =\frac{2\pi }{\beta }k\, .\end{equation}
Using the following symmetries of the folded sine-Gordon model
\begin{equation}
\Phi\,\rightarrow\,\Phi+\frac{2\pi}{\beta}\quad ,\quad \Phi\,\rightarrow\,-\Phi\,,
\end{equation}
it can be easily shown \cite{kfold} that
\begin{equation}
\label{nth_vev}
\mathcal{G}^{(k)}_{a}(\beta )=\big \langle k\big |V_{a}(0)\big |k\big \rangle =\mathcal{G}_{a}(\beta )e^{i\frac{2\pi a}{\beta }k}\: .
\end{equation}
The form factors corresponding to states above these vacua can be
obtained by substituting $ \mathcal{G}_{a}(\beta ) $ with $
\mathcal{G}^{(k)}_{a}(\beta ) $ in the form factor formulae
(\ref{ff_11}, \ref{ff_1111}, \ref{ff_22}) above.


\begin{thebibliography}{1}
\bibitem{delfino_mussardo}G. Delfino and G. Mussardo, \emph{Nucl. Phys.} \textbf{B516} (1998) 675-703,
hep-th/9709028.
\bibitem{nersesyan}M. Fabrizio, A. O. Gogolin and A. A. Nersesyan,
\emph{Nucl. Phys.} \textbf{B580} (2000) 647-687, cond-mat/0001227.
\bibitem{bullough} R.K. Bullough, P.J. Caudrey and H.M. Gibbs, in
\emph{Solitons}, Eds. R.K. Bullough and  P.J. Caudrey, Topics in
Current Physics v. 17, Springer--Verlag, 1980.
\bibitem{mass_scale}Al. B. Zamolodchikov, \textit{Int. J. Mod. Phys.} \textbf{A10} (1995) 1125-1150. 
\bibitem{yurov_zamolodchikov}V. P. Yurov and Al. B. Zamolodchikov, \emph{Int. J. Mod. Phys.} \textbf{A6}
(1991) 4557-4578. 
\bibitem{frt2}G. Feverati, F. Ravanini and G. Tak{\'a}cs, \emph{Phys. Lett.} \textbf{B430}
(1998) 264-273, hep-th/9803104.\\
G. Feverati, F. Ravanini and G. Tak{\'a}cs, \emph{Nucl. Phys.} \textbf{B540}
(1999) 543-586, hep-th/9805117.
\bibitem{kfold}Z. Bajnok, L.Palla, G. Tak{\'a}cs and F. W{\'a}gner: \emph{The k-folded sine-Gordon
model in finite volume,} ITP-BUDAPEST-557, KCL-MTH-00-21,
hep-th/0004181, to appear in \textit{Nucl. Phys.} \textbf{B}. 
\bibitem{klassen_melzer2}T.R. Klassen and E. Melzer, \emph{Nucl. Phys.}
\textbf{B370} (1992) 511-550. 
\bibitem{nonintegrable}G. Delfino, G. Mussardo and P. Simonetti, \emph{Nucl. Phys.} \textbf{B473} (1996)
469-508, hep-th/9603011. 
\bibitem{luscher} M. L{\"u}scher in \emph{Champs, cordes et phenomenes
critiques}, Proc. Les Houches Summer School, ed. E. Brezin and
J. Zinn-Justin (North-Holland, Amsterdam, 1989).
\bibitem{klassen_melzer1}T.R. Klassen and E. Melzer, \emph{Nucl. Phys.}
\textbf{B350} (1991) 635-689. 
\bibitem{exact_vevs}S. Lukyanov and A.B. Zamolodchikov, \emph{Nucl. Phys.} \textbf{B493} (1997)
571-587, hep-th/9611238. 
\bibitem{formfactors}S. Lukyanov, \emph{Mod. Phys. Lett.} \textbf{A12} (1997) 2543-2550, hep-th/9703190. 
\end{thebibliography}
\end{document}